\documentclass[a4paper,11pt]{article}
\usepackage{pos}
\usepackage{adjustbox}
\title{Dusty Dark Matter Bubbles of a New Vacuum  stopped in
  Earth
and Radiating 3.5 keV X-rays.}
 \ShortTitle{Dusty Dark Matter Bubbles, X-rays}
\author*[a]{Holger B. Nielsen}
\author[b]{Colin D. Froggatt}
\affiliation[a]{The Niels Bohr Institute,\\ Blegdamsvej  15 - 21, Copenhagen,
  Denmark}
\affiliation[b]{Physics and Astronomy, Glasgow University,\\ Glasgow G12 8QQ, Scotland}
\emailAdd{hbech@nbi.dk}
\emailAdd{Colin.Froggatt@glasgow.ac.uk}
\abstract{We continue our work on a proposal for what dark matter could be,
  namely that the dark matter consists of essentially macorscopic objects built from
  ordinary matter. The only element of ``new physics'' is that there should
  exist several types or phases of vacuum. Then the dark matter particles
  are   bubbles of a new type of vacuum filled with ordinary matter
  - say diamond -
  under high pressure and the whole bubble is supposedly contaminated by the
  materials in the extragalactic or intragalactic space able to form dust.
  The dust has a lower than dimension 3 structure and may well be long chains of atoms.
  
  We begin speculations on identifying the phase transition between the vacua, which we require in our model, with a
  (second order?) transition revealed in the Columbia plot with quark masses as axes. In the corner of small quark masses there is a true phase transition as temperature is raised, while outside this region there is instead no genuine phase transition under variation of the temperature but rather only a  “crossover”.}
\FullConference{Workshop on the Standard Model and Beyond, Part of
  Corfu Summer Institute,
  22nd Hellenic School and Workshops on Elementary Particle Physics and
  Gravity, Corfu, Greece 2022\\
Corfu\\
28th August to 8th September
}

\begin{document}


\maketitle


  \titlepage
  \section{Introduction}

  Concerning its gravitational force the existence of dark matter is
  incorporated in the standard cosmological model, but the non-gravitational
  properties of the dark matter is much less well-established and even connected
  with seeming contradictions or mysteries.
  
   Many physicists still believe in WIMPs being the dark matter, while
  we ourselves go for a model hopefully incorporable into the Standard Model, in which the dark matter is more like macroscopic structures of ordinary matter (only with one new vacuum story added). But let us start by looking at a few questions:
  
  \subsection{ Mysteries concerning Dark Matter}
\begin{figure}
  	\includegraphics[scale=1.3]
  	{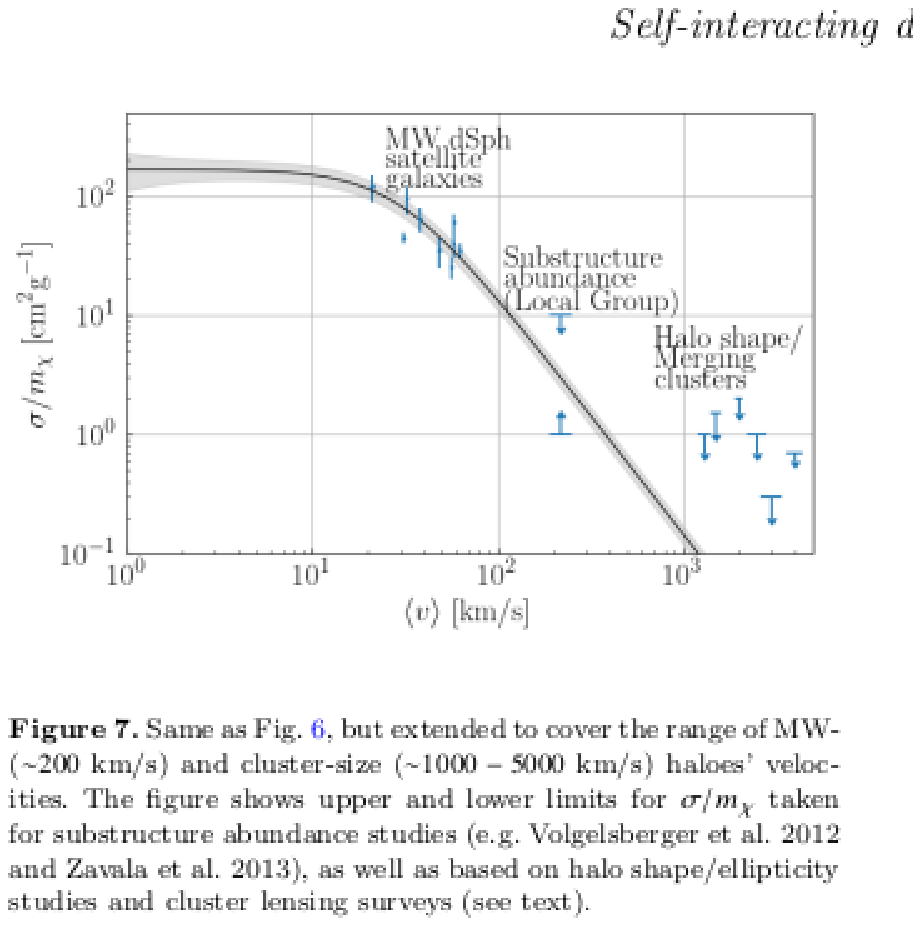}
  	\caption{By fitting to various dwarf galaxies Correa \cite{CAC} has
  		obtained the ratio of the cross section for one dark particle
  		hitting another one divided by the mass of such a particle,
  		$\frac{\sigma}{M}$, also called the inverse darkness, as a function
  		of the velocity in the collision. Since varying the mass and the
  		cross section by the same factor would not change any effect of the
  		dark matter, one can of course never obtain anything but this ratio from the dark matter effects on the ordinary matter.}
  	\label{Correa}
  \end{figure}

  \begin{itemize}
  \item {\color{red} Why do the Xenon experiments {\em not} ``see'' the dark
    matter ?}
    
\vspace{1 mm}
       {\color{blue} The answer of our model:} The dark matter particles get
       stopped down to
       too low speed by the atmosphere and the shielding.  Then they do not have enough speed to bring the nuclei in the detector
       up to a speed that gives visible scintillation in the counters.
       
  \item {\color{red} But how can then DAMA ``see'' the dark matter?}
    
\vspace{1 mm}
       {\color{blue}Our answer:} Dark matter radiates electrons. DAMA only tests that it
       is dark matter
       by the seasonal variation \cite{DAMA1,DAMA2}, while the Xenon-experiments exclude (or can
       separate) the electron
       induced signal. So DAMA would accept as possible dark matter a signal
       with sufficiently high energy electrons. However such high energy electrons would not be produced by simple collisions of a few hundred
       km/s dark matter particles; so special radiation would be needed.
       
  \item {\color{red} Could the Xenon experiments then not look for electrons
    also?}

    \vspace{1 mm}
           {\color{blue} Our answer:} Yes, and indeed Xenon1T ``saw'' a little
           excess of
           ``electron recoil events'' of energy about 3.5 keV \cite{Xe1T}.

           Unfortunately this little electron-recoil effect  was
           retested and it turned out, as reported in a paper published shortly after this workshop \cite{XenT}, to probably have been due 
           to the presence of radon gas. So now the situation is
           that the Xenon-experiments saw nothing in electron recoil events either.
    

  \item {\color{red} But recoil electrons from 300 km/s collisions only
    have about 1 eV energy, not keV's!}

    \vspace {1 mm}
            {\color{blue} Our model answer:} Our dark matter particles can
            be excited so as to radiate 3.5 keV X-rays or electrons.

          \item {\color{red} How does the dark matter get excited so as
            to radiate electrons in DAMA?}

            \vspace{1 mm}
                   {\color{blue} Our answer:} The dark matter particles get
                   heated up and excited by the interaction with the atmosphere
                   while being stopped.
    
                 \item {\color{red} But if they get so easily stopped and
                   slowed down, does it not mean that the dark matter is not
                   truly dark?}

                   \vspace{1 mm}
                          {\color{blue} Answer:} It has an ``inverse darkness''
                          \begin{eqnarray}
                            \frac{\sigma}{M} &=& 15 m^2/kg \hbox{ (for
                              velocity $v\rightarrow 0.$ )}  
                          \end{eqnarray}
                          enough for stopping, but only little noticed in
                          astronomy.

                          In fact Correa made a fit of dwarf galaxy star
                          velocities in a model with self interacting dark matter,
                          SIDM, and obtained the value $15m^2/kg$ for the low
                          velocity limit of the ratio of the cross section
                          relative to the mass of a dark matter particle \cite{CAC}.
                          At higher velocities the cross section falls off. Her results are displayed in Figure \ref{Correa}.
    \end{itemize}
  
\subsection{Our only moderately dark  Dark Matter}

In our model the dark matter particles or pearls consist of a tiny bubble of a new speculated vacuum filled with highly compressed ordinary matter, say carbon, surrounded by an approximately 100 times larger dust grain. This dust grain can be washed off in the Earth's atmosphere, leaving behind a cleaned vacuum bubble with a cross section say 10000 times smaller than that of the dusty pearl. The cleaned dark matter pearl gets decelerated to a much slower speed in the atmosphere and can then penetrate slowly through the shielding of the earth down to say DAMA. 
  

 
Thus we have for the dusty pearls (at low velocity and at 1000 km/s in big galaxy clusters) and cleaned pearls the following values for the "inverse darkness '' $\frac{\sigma}{M}$ (where $\sigma$ is the cross section of the pearl and $M$ its mass): 




   \begin{eqnarray}
    \frac{\sigma}{M}|_{v\rightarrow 0} &=& 15 m^2/kg \hbox{ (``observed'', dusty)}\\
    \frac{\sigma}{M}|_{clean} &=& 10^{-3}m^2/kg \hbox{ (cleaned, vacuum-bubble)}\\
    \frac{\sigma}{M}|_{v=1000km/s}&=& 10^{-2} m^2/kg \hbox{ (``observed'', dusty, big clusters)}
    \end{eqnarray}
    
   Note that even our dusty dark matter is dark compared to atoms, but conventional WIMP dark matter is much darker than ours:  
  \begin{eqnarray}
    \hbox{Carbon C: }\frac{\sigma}{M}|_{C} &=& \frac{\sigma_C}{12u*1.66*10^{-27}kg/u}=7.73 *10^5m^2/kg\nonumber\\
    \hbox{WIMP say: } \frac{m_W^{-2}}{m_W}&\sim &  \frac{(0.2*10^{-17})^2m^2}
         {2*10^{-27}kg} \sim  10^{-8}m^2/kg \nonumber  
  \end{eqnarray}
 \begin{figure}
	\includegraphics
	{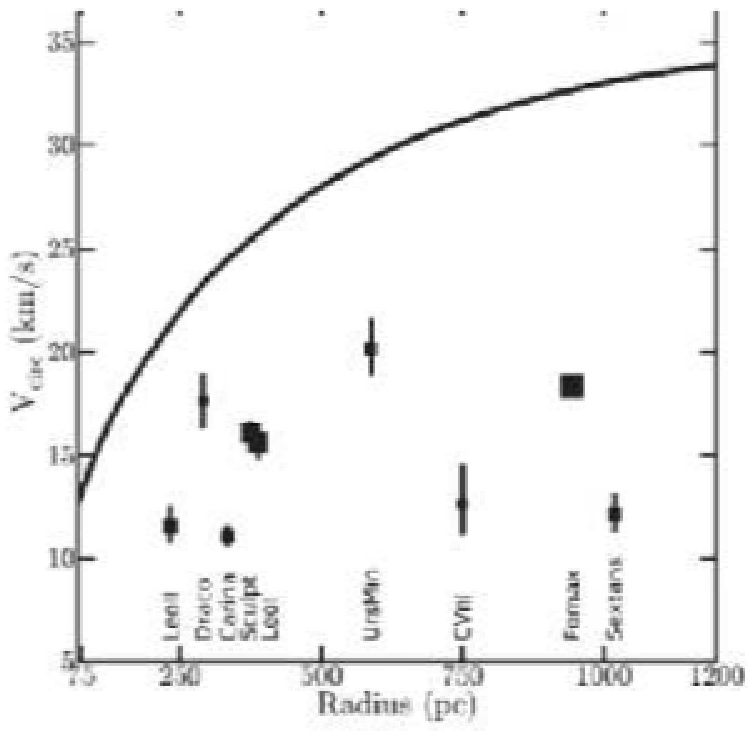}
	\caption{This figure should illustrate that when Correa
		simulated the dark matter distribution, under the assumption of a purely gravity interaction, and regained a prediction
		for the star velocities it was {\em not} successful. So there is a call
		for e.g. interaction of the dark matter with itself.}
\label{gravitation only}	
\end{figure}
   


  Figure \ref{gravitation only} shows that  dark matter with only gravitational interactions does not function so well in  simulations on dwarf galaxies \cite{CAC}.     

  


   \section{ Our Model of Dark Matter in principle inside the Standard Model}
   Provided we can invent some theoretical story like that QCD, instead of
   being simply confined and having the Gellman $SU(3)$ symmetry of rotating the light quarks into each other or better the chiral part of it being
   spontaneously broken, could have another phase where the diagonal
   subgroup of the two $SU(3)$s survives spontaneous breaking \cite{Wilczek}. 
   Then we could speculate on the existence of two phases of the QCD-part of the vacuum-structure.
   Really QCD and the usual Nambu-JonaLasinio spontaneous breaking is
   sufficiently complicated that a slightly {\em different QCD-part} of the
   vacuum structure can {\em not be excluded}. If such a possibility had just been overlooked so far then, due to the QCD-physics, there could be two phases of the vacuum,
   and our speculation about an extra vacuum inside of the dark
   matter bubbles could be realized completely inside the Standard Model.
   This could of course explain why LHC did not see any new physics in
   spite of it usually being strongly expected from considering dark matter.
   
   \begin{itemize}
   \item{\color{red} Why did LHC not see any dark matter?}

     \vspace{1 mm}
            {\color{blue} Our answer:} In our model the dark matter is composed
            from ordinary stuff like nuclei and electrons being caught into
            a bubble of a second phase of the vacuum.

            There is no ``new physics'' in our model, except that we do not
            know how the two (or more) types of vacua come about.

            But we speculate that they can appear inside the Standard Model
            without genuine new physics; only with fine tunings of couplings.

            Presumably: {\em One vacuum with confinement; another one  with QCD
              color Higgsed to be aligned with a Gellman $SU(3)$. Another idea is that Nambu-JonaLasinio Goldstone boson fields (the pseudoscalar meson fields) could be correlated a little differently in the different vacuum-phases.} We shall show some ``Columbia-plots'' at the end that might give
              hope for this latter ``different correlations in Nambu-JonaLasinio''
              idea.
                        
    \end{itemize}
 There is another difficult question
 \begin{itemize}
 	\item{\color{red} If DAMA can observe dark matter particles emitting fast electrons with energies of the order of 3.5 keV, then why can the Xenon-experiments not see them?}
 	
 	\vspace{1 mm}
 	{\color{blue} Our model answer:} {\em This is because the Xenon experiments have a fluid and when the dark matter particles come to the fluid they fall much faster than in the solid NaI in the DAMA experiment.} So the dark matter particles spend much less time in one kg of xenon than in one kg of NaI. 
 	
 	This problem is very hard to solve with a WIMP-type model except if mysteriously the nuclear physics
    is very different for xenon versus the other elements Na and I, which sounds almost like
 	fine-tuning the xenon interaction with the WIMP to have a special cancellation!
 	The fluid solid difference between the two scintillators is the almost only proper qualitative
 	difference between them. (Remember that LUX and other xenon experiment have looked so
 	accurately that had there even been the amount of WIMPs that would give the seasonal excess in
 	DAMA, they should have been able to see it.)
 
  A related question is:
\item{\color{red} Is there a way to check our dark matter model?}
\vspace{1mm}

	{\color{blue} Our answer:}
	One good check of our model would be to confirm that the DAMA signal for dark matter is due to the emission of fast electrons rather than nuclear recoils. In fact the COSINUS experiment \cite{Cosinius} is presently being set up in Gran Sasso using NaI scintillating crystals as a target to cross-check the DAMA results. It is capable of disentangling $e^-/\gamma$ events from nuclear recoils on an event-by-event basis, which should provide a good test of our model. 

\end{itemize}

   \section{\label{Pearl} Our picture of the Dark Matter Pearls}
   In our model \cite{Dark1,Dark2,Tunguska,supernova,Corfu17,Corfu19,theline,Bled20,Bled21,extension,Corfu21,Bled22} the dark matter consists of macroscopic massive pearls, which
   in the interior has a little piece - very exactly a perfect sphere - of
   a vacuum in phase 2 so to speak filled with some ordinary material, such as say carbon, under high pressure. Then this little bubble of new vacuum
   is surrounded by a bigger but not so heavy grain of dust. The point is that,
   even though the ordinary material inside the bubble is in the new vacuum
   and highly compressed, it still interacts chemically and mechanically with
   the dust and thus presumably does not stay clean in the Universe for 13.6
   milliards years even if it should have been created clean, in the sense that
   at first there was plasma around it rather than material that could
   contaminate it.

  The dark matter particles or pearls are composed of:
    \begin{itemize}
    \item A nm-size bubble of a new speculated vacuum filled with
      highly compressed atomic stuff, say carbon. (We really have only
      very weak constraints on the size of the pearls.)
      
      Let us remark that there is not so much to observe that can give us a good estimate of the mass of the dark matter particles. Therefore our estimate of the size of the dark matter pearls is very uncertain. If we use the idea to be found in our talk on "Tensions in Cosmology'' \cite{walls} that the cosmological constant might be replaced by a network of domain walls, a surface tension of the dark matter bubbles of the order of (30 MeV)$^3$ would be called for. Then dark matter bubbles would be 
      of $10^{-8}m$ size, an order of magnitude 
      larger than the here mentioned nanometre size. But, as stressed above, the mass is not well determined. 
      
    \item A surrounding dust particle of ``metallicity'' material
      such as C, O, Si, Fe, ..., presumably of some non-integer
      Hausdorff dimension about 2. This atomic matter is influenced
      by the electrons being in a superposition partly inside
      the bubble of the new vacuum, where there is a very high gap
      between filled and unfilled electron states.
    \end{itemize}
   The pearls interact:
    \begin{itemize}
    \item with other dark matter particles,
    \item with atomic matter.
      \end{itemize}

   \begin{figure}
  	\includegraphics[scale=0.6]
  	{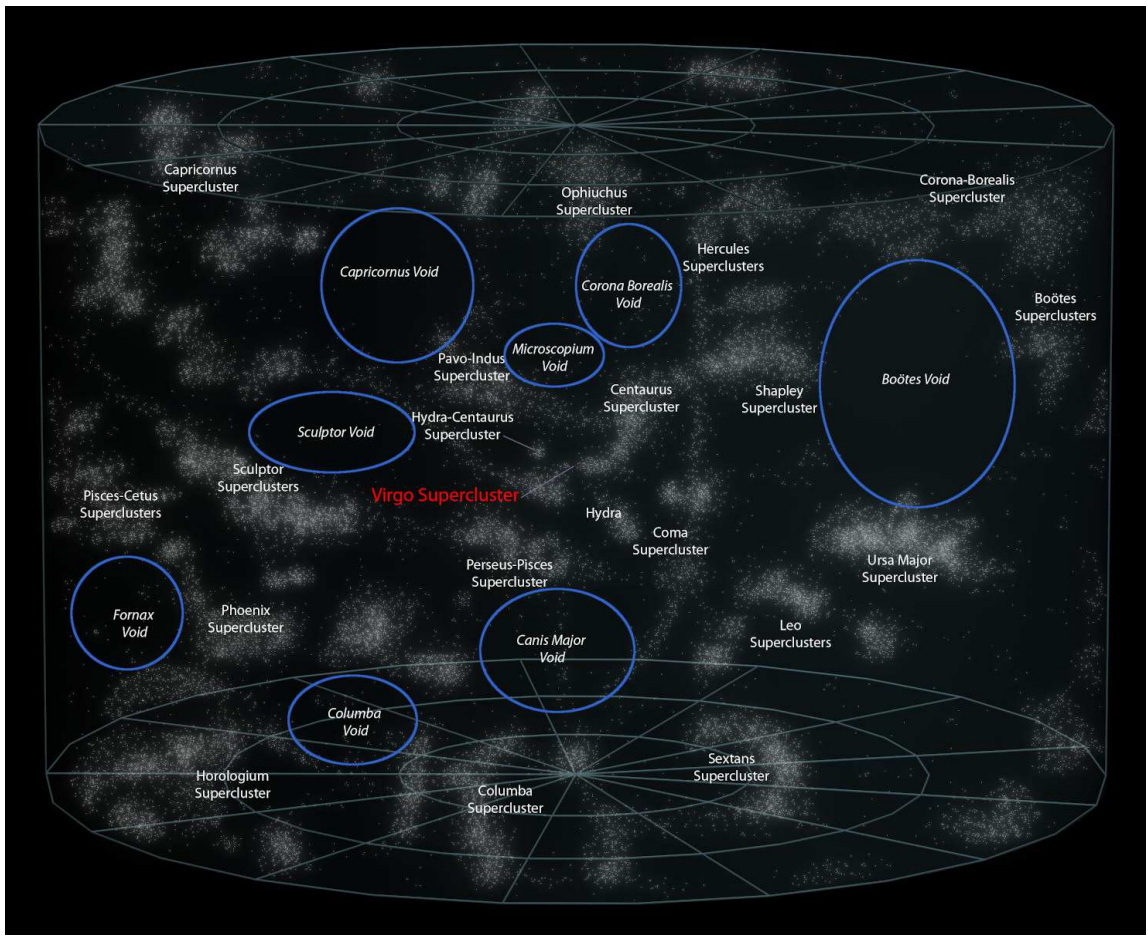}
  	\caption{This picture (from Wikipedia) is an attempt to draw 
  		the distribution of galaxies in perspective as 3-dimensional. It shows several voids, regions
  		with only a few galaxies, and by accident has rings around these
  		voids. You could roughly take these rings to suggestively represent
  		domain walls, in a speculation that the voids were of a different
  		type (phase)
  		of vacuum than the galaxy rich regions (clusters etc.).} 
  	\label{Voids}
  \end{figure}
  
  \section{Huge New Vacuum Bubbles possible, if really there are different
    Vacua}

  If really the idea - our main new idea - that there are several
  phases of vacuum were true, as well as our assumption that the different
  phases have the same energy density, then a priori one could expect huge,
  astronomically large, regions with a different vacuum phase from that of another large region. If so, however, there would still be the
  surfaces - the domain walls - in between these different regions.

  Now if these surfaces in between the regions would have energy per area
  densities as expected by dimensional arguments for new physics beyond
  the Standard Model, then these densities of energy on the domain wall
  would be so huge that astronomical size domain walls would upset
  our usual cosmological picture very strongly. Only because 
  the domain wall energy density needed to fit our dark matter turns
  out to be surprisingly small from the new physics expectation point of view, does it become possible to speculate seriously that the domain
  walls could really exist out in space - and e.g. surround the
  voids (the big regions in space with rather few galaxies) - and only contribute tolerably to the cosmological parameters. This is illustrated in Figure \ref{Voids}.

 
  
  The idea of having huge astronomical size vacuum bubbles will be taken up
  in our contribution to the Corfu workshop  on ``Tensions in Cosmology'' just after this workshop \cite{walls}.
  
  \section{Multiple Point (Criticality) Principle}
  
  As mentioned above our main new idea which we incorporate into the Standard Model is the Multiple Point Principle (MPP)  \cite{MPP1,MPP2,MPP3,MPP4}, according to which there should be several phases of the vacuum and that they should be degenerate in the sense of having the same energy density. This principle can be used to fine-tune and hence predict the value of coupling constants. It was applied some time ago at the Planck scale in a somewhat complicated model to correctly predict the number of families in the Standard Model \cite{Picek, book}, prior to the LEP  measurement of the number of light neutrino species. 
  
\begin{figure}
	\includegraphics[scale=0.55]{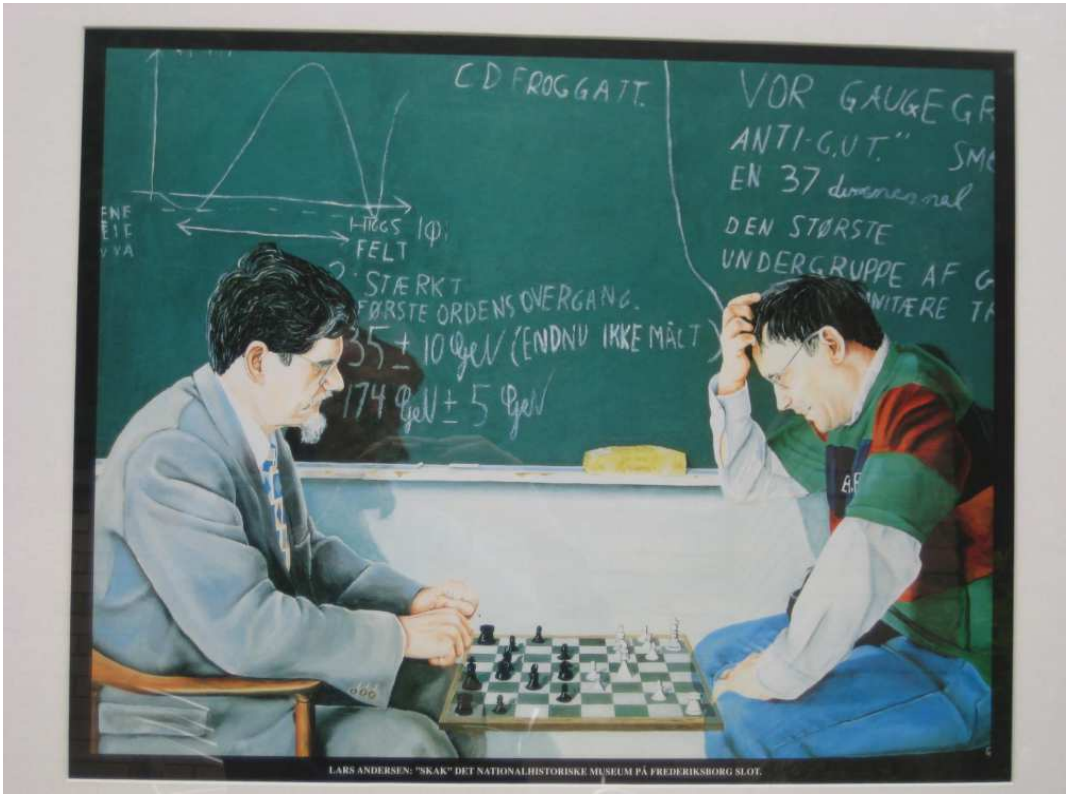}
	\caption{ This painting of one of the authors and the Danish
		finance minister was painted long before the Higgs particle
		was observed at LHC. Nevertheless due to our ``Multiple Point
		Criticality Principle'', we predicted the mass value for the
		Higgs particle $ 135 \, GeV \pm 10\, GeV$ as with the $1$ hidden by
		Mogens Lykketoft's head is seen on the painting. The much later
		measured value is $125\, GeV$.} 
	\label{Painting}
\end{figure}

  Later we used MPP
  to PREdict the mass of the Higgs particle before this particle was found experimentally \cite{tophiggs,Corfu1995}. In addition to the usual vacuum with a Higgs vacuum expectation value of 246 GeV, we obtained another vacuum degenerate with it but having  a very large Higgs field expectation value of order the Planck scale $\sim 10^{18}$ GeV. In Figure \ref{Painting} we reproduce a copy of a painting including one of us, in which  
  the Higgs mass prediction $135\, GeV \pm 10\, GeV $ is on the black board
  (although the $1$ in the $ 135$ is hidden behind the member of the cabinet
  Mogens Lykketoft's head).

\begin{figure}               
\includegraphics[scale=1.2]{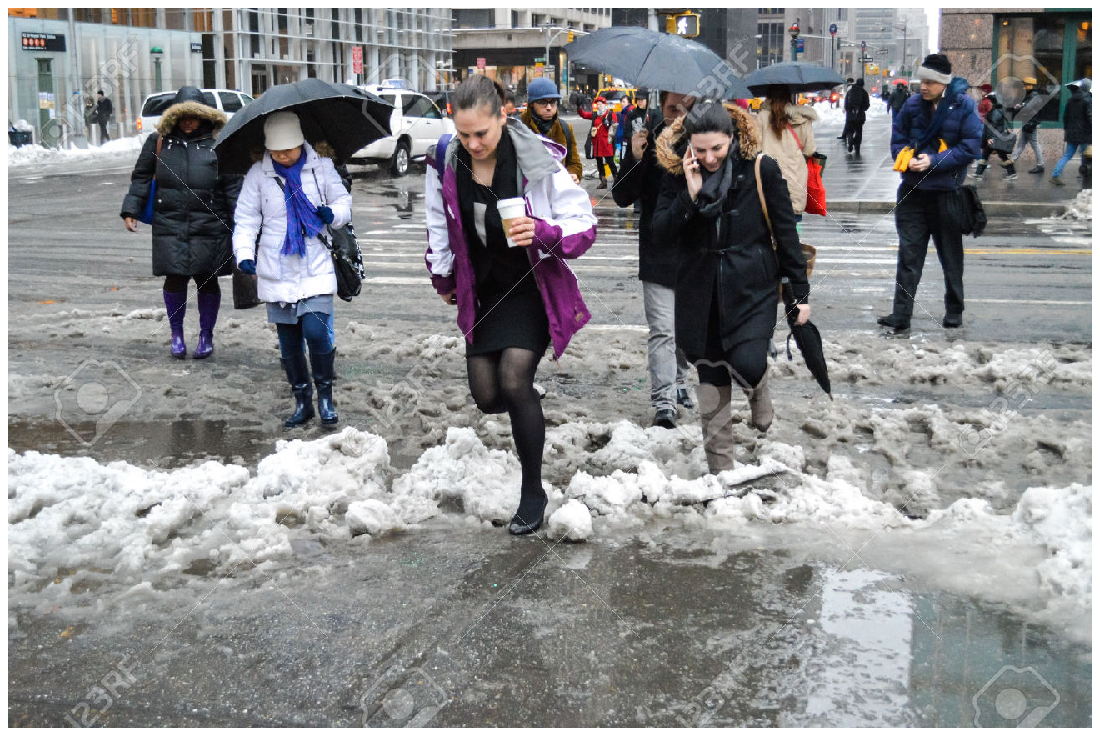}
\caption{Slush.}
\label{Slush}
\end{figure}

                There may not exist any true derivation of our ``Multiple
                Point Criticality
                Principle'', because it is basically an assumption.
                A good argument for it is very similar to the argument
                for having the melting point temperature when you have
                both say ice and water together. It occurs often and even
                has a name ``slush'', and having slush one knows that the
                temperature is $0^0$ Celsius (see Figure \ref{Slush}).

\section{
  The 3.5 keV X-ray from Dark Matter
}

    The most important support for our model may be that we find the
    energy 3.5 keV in {\em two} different places 
    as a possible characteristic
    level transition energy difference for dark matter:
    \begin{itemize}
    \item The energy of the X-ray line from galaxy clusters, galaxies and 
    strangely the Tycho supernova remnant long suspected as coming from dark matter.
    \item As average of the energy of the DAMA dark matter events.
    \end{itemize}
%
    	
    	In section \ref{addendum} we report on some progress made since the workshop on our understanding  
    of the observed intensity of the 3.5 keV X-ray radiation. Our picture
    now is that most of the sources are well understood by the fit of
    Cline and Frey \cite{Cline}, in which the rate of this line emission goes as the square of the dark matter density. This corresponds to the dominating process being a collision or perhaps annihilation of two dark matter particles, 
    which get some excitation emitting the 3.5 keV line. However, of course
    the emission in a visible amount from the Tycho supernova remnant cannot be
    dominantly due to this dark + dark collision \cite{Jeltema}, but must involve some
    ordinary matter colliding with the dark matter. Also the outskirts of the
    Perseus Cluster, where the dark matter density has fallen, and the centre
    of the Milky Way are two further sources of the 3.5 keV line,
    where we now think that it is collision of ordinary matter - expected to be
    hot atoms with energy of the order of X-rays - with the dark matter that
    excites the latter to emit the 3.5 keV line.

    In our model we have long supposed that the dark matter bubbles of new
    vacuum contain matter in which the electron spectrum has a gap, a homolumo
    gap between the filled levels and the empty levels. It is this homolumo
    gap which, because of the very high pressure in the ordinary matter
    contained inside the bubble, has the unusually high value (compared to ordinary chemistry) of 3.5 keV for the energy difference between the highest(H) occupied(O) level (molecular orbit = MO) and the lowest(L) unoccupied(U) molecular orbit. This should then explain the tendency for the dark matter bubble to emit an X-ray of just this frequency.
    The excited electrons should go to the lowest unoccupied states and from there jump, under 3.5 keV X-ray line emission, down to a hole in the band of normally occupied states.

    \section{The Dust sitting on the Dark Matter Pearls/Bubbles}

    Since our bubbles of new vacuum contain - are pumped up with -
    ordinary matter (under pressure) it is not surprising, but
    rather expected, that a chemical interaction could be formed 
    between the ordinary matter inside the bubble and the ordinary matter
    outside. In fact we expect that, due to the high pressure, some electrons
    will already by the Fermi statistics flow a bit over and actually
    be present outside the proper skin (= domain wall). So the bubbles have a
    high chance of getting contaminated by dust or other matter in the outer space where they move along. So we should imagine that the true dark matter
    particles are probably much like dust grains with a bubble of new vacuum
    filled with say carbon sitting in it somewhere. Or, if the bubbles are very
    large, they may literally be dirty bubbles. 
    
\begin{figure}
      \includegraphics[scale=7]{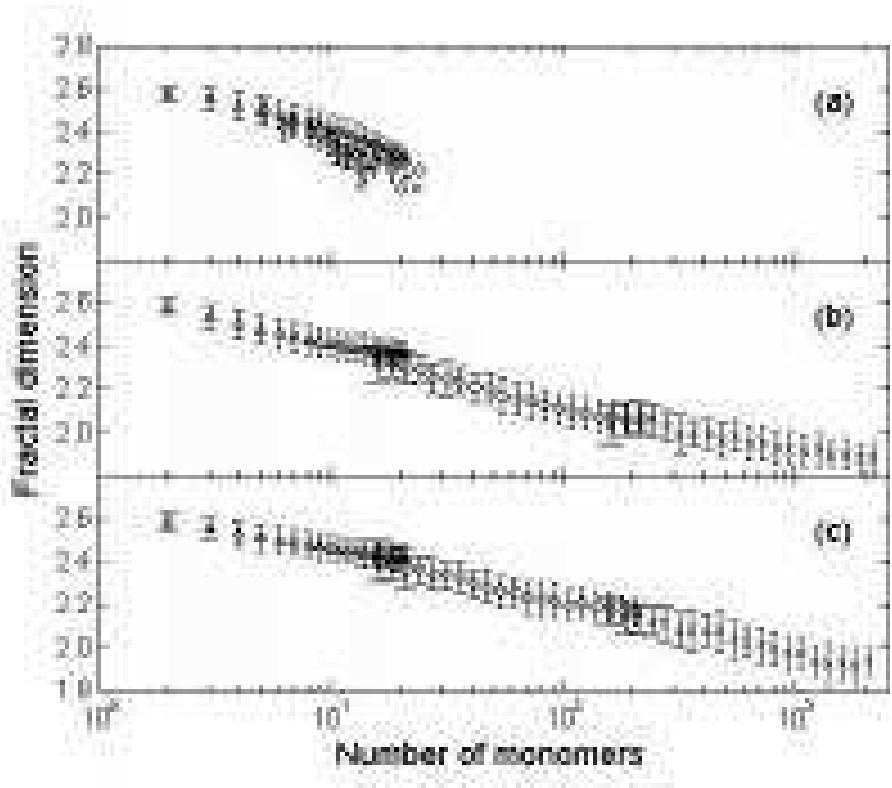}
    \caption{ Results of simulations of dust grains growing by adding single atoms or molecules \cite{Hd} suggest that
    	the Hausdorff dimension can become low of the order 1 or 2.}
    \label{Hayes}
\end{figure}
\begin{figure}
      \includegraphics[scale=0.5]
          {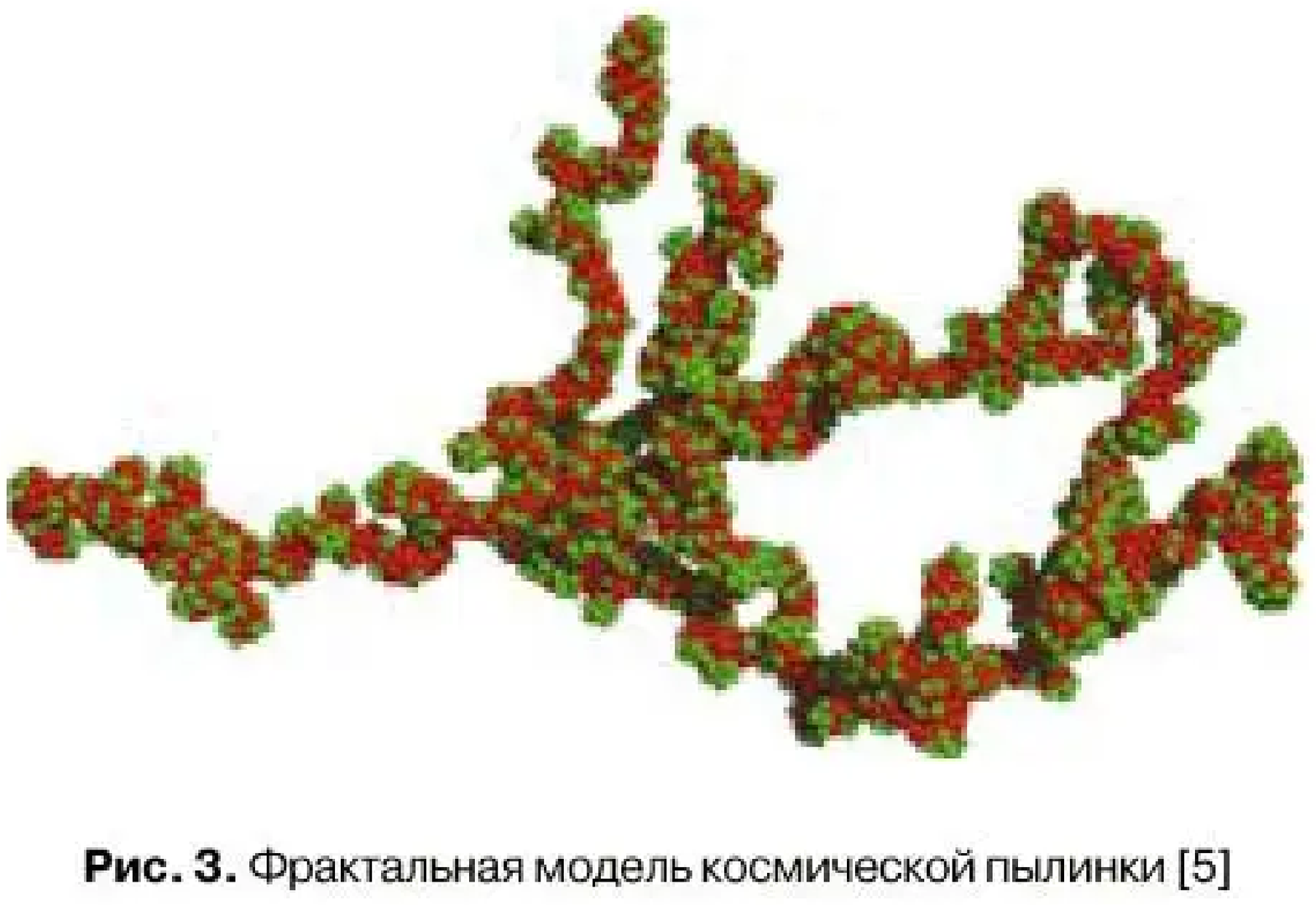}
          \caption{Picture of a fractal cosmic dust grain constructed from 1024 monomers \cite{Wright}. The mentioned lower Hausdorff dimensionality of the dust grains - if say one dimensional - would mean that
          	the dust grain is really more like some knot of strings of molecules than a pearl.}
          \label{dust}
        \end{figure}

    Simulations of the growth of a dust grain, in which the atoms get attached to the dirt rather few at a
    time \cite{Hd}, show that the dust grain easily develops to have a lower Hausdorff dimension than a compact dust ball would have. The dust could easily be more like a little surface or some stringy network structure. 
    The growth of a dust grain and the development of its fractal dimension in such a simulation is given in Figure \ref{Hayes} .
An example of such a fractal cosmic dust grain built up from 1024 monomers \cite{Wright} is given in Figure \ref{dust}.

      
\begin {figure} 
      \includegraphics[scale=0.5]
          {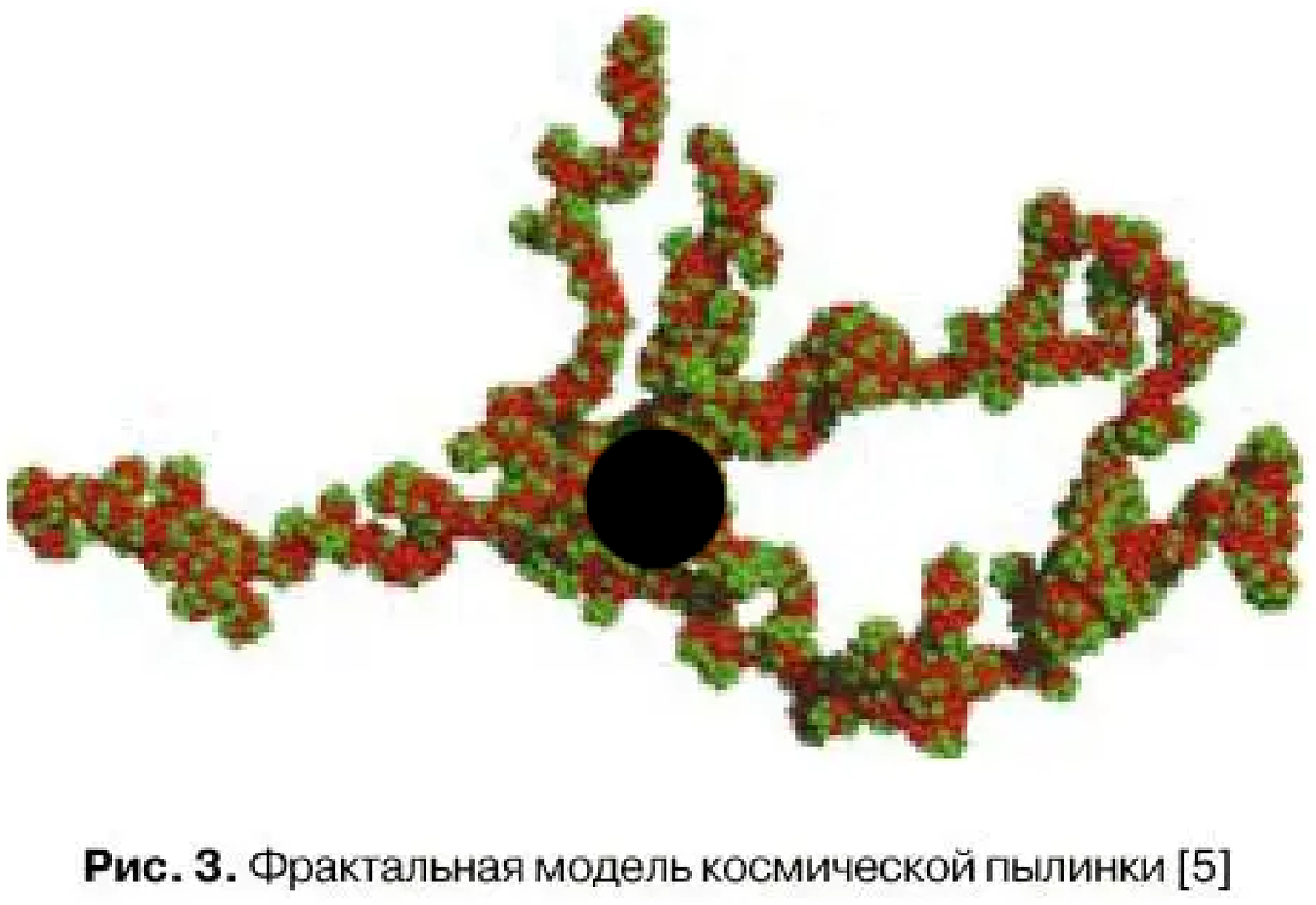}
          \caption{Our dark matter particle is a bubble with a dust grain round it. The bubble is supposedly much heavier than the dust grain.}
          \label{dustypearl}
\end{figure}

      We expect that such a dust grain would collect on top of one of our dark matter pearls, which in itself is very much like a seed atom. We may illustrate that in Figure \ref{dustypearl} by drawing our little pearl as a bubble of new vacuum inside the dust grain.
      If the grain size is $0.1 \mu m$ and the bubble size is $1 nm = 0.001 \mu m$, then
      the bubble is about 100 times smaller than the dust grain. But now if we should seek to implement the ideas of adjusting the tension in
      the domain walls to make it possible for the domain walls to take over the role of the cosmological
      constant, we should take somewhat larger bubbles in the middle and the dust grain would not be so much
      larger than the bubble of the second vacuum.
    
\section{Achievements}
      Important achievements of our model:
      \begin{itemize}
      \item Explains that only DAMA ``see'' the dark matter by the
        particles interacting so strongly as to be quite slow and unable to
        knock nuclei hard enough to make observable signals. Instead
        the DAMA signal is explained as due to  emission of electrons 
        with the ``remarkable energy of 3.5 keV'' from pearls in an excited state.

      \item The characteristic frequency of electron or photon emission
        by the dark matter particles is due to a homolumo gap in the
        material inside the bubble of the new vacuum. This gap should be
        equal to the 3.5 keV. 
        \end{itemize}

       \begin{figure}
      	\includegraphics
      	{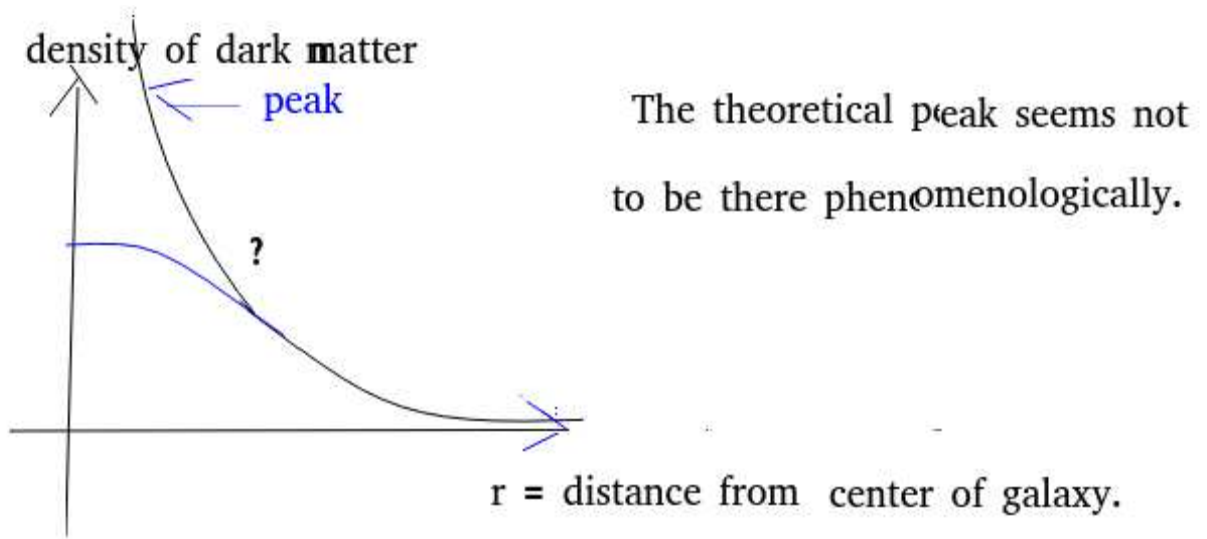}
      	\caption{Illustration stressing that the pole peak in the ideal NFW-distribution at the centre of a galaxy, which is what truly only
      		gravitationally interacting dark matter should form, is phenomenologically not quite as sharp as it should be. This could reveal self interaction of dark matter SIDM as in Correa's fits.}
      	\label{NFW2}
      \end{figure}
      
      \begin{itemize}
      \item We have made a rather complicated calculation of what happens when the
        bubbles - making up the main part of the dark matter particle - hit
        each other and the surface/skin/domain wall contracts and how one gets
        out a part of the energy as 3.5 keV X-rays \cite{theline}. We fit with one parameter both the very frequency 3.5 keV, and the
        overall intensity of the corresponding X-ray line observed from galaxy clusters etc. The production mechanism gives an intensity proportional to the dark matter density squared and we use the results of the analysis of Cline and Frey \cite{Cline} whose model shares this property.
      \item We explain why - otherwise mysteriously - the 3.5 keV line
        was seen by Jeltema and Profumo \cite{Jeltema} from the Tycho supernova remnant and probably also explain problems with the Perseus galaxy cluster 3.5 keV
        observations. This is by claiming the excitation of the bubbles
        comes from cosmic radiation in the supernova remnant.  After the workshop we now tend to think that
        it is not the cosmic rays, but rather hot atoms that are most important. Cosmic rays would
        survive so long that they would make embarrassingly high intensities for the 3.5 keV line from say the
        supernovae in the Milky Way Center.
        \end{itemize}

      
      \begin{itemize}
      \item According to expectations from ideal dark matter that only
        interacts essentially by gravity, there should be e.g in a dwarf galaxy
        a concentrated peak or cusp of dark matter (the NFW distribution \cite{NFW}), but that seems not to be true.
        The inner density profile rather seems to be flat as expected for self-interacting dark matter \cite{firstSIDM}, as illustrated in Figure \ref{NFW2}.
        Correa \cite{CAC} can fit the dwarf galaxy star velocities by the hypothesis that dark matter particles interact with each other with a cross section
        over mass ratio increasing for lower velocity, as shown in Figure \ref{Correa}. We fit the cross section
        over mass velocity dependence of hers. But we need a ``hardening ''
        of the dust around the bubbles. 
        
        We have already seen in Figure \ref{Correa} how
        the inverse darkness $\frac{\sigma}{M}$ of the dark matter varies with velocity roughly in the way that it stands at about
        $15m^2/kg$
        up to velocities of the order of 220 km/s. For higher velocity it begins to fall 
        and becomes so small that one mainly has upper bounds for  $\frac{\sigma}{M}$ for velocities present in the interior
        of big galactic clusters.
      \end{itemize}
      

      
  
 \begin{figure}
	\includegraphics[scale=0.7]
	{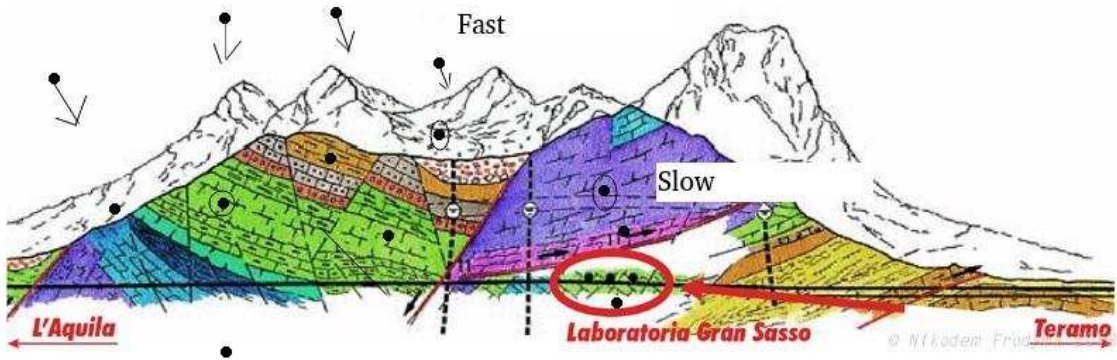}
	\caption{ The mountains above the Gran Sasso laboratories.}
	\label{ip}
\end{figure}
    
    \section{Impact}
    
    

  
      We imagine that the dark matter pearls lose their dust grain in the
      atmosphere or at least, if not before, by the penetration into the earth
      shielding and that they at the same time get excited by means of the
      energy of the braking of the pearls. For a very small number of the pearls this
      excitation energy gets radiated first much
      later when the pearl has passed through the earth shielding to the underground detectors (see Figure \ref{ip}), so as to
      deliver X-ray radiation with just the characteristic 3.5 keV energy
      per photon. The energy is delivered we guess by electrons or photons.
      Thus experiments like the xenon experiments do not ``see'' it
      when looking for nucleus-caused events. Only DAMA, which does
      not notice if it is from nuclei or from electrons, does not throw
      electron-caused events away as something else.

  
  The dark matter pearls come in with high speed (galactic velocity), but
  get stopped down to a much lower speed by interaction with the atmosphere and the shielding
  mountains, whereby they also get excited to emit 3.5 keV X-rays or
  {\em electrons}.
  \subsection{The Xenon experiments versus DAMA Contradiction}
 
\begin{figure}
 	\includegraphics[scale=0.75]{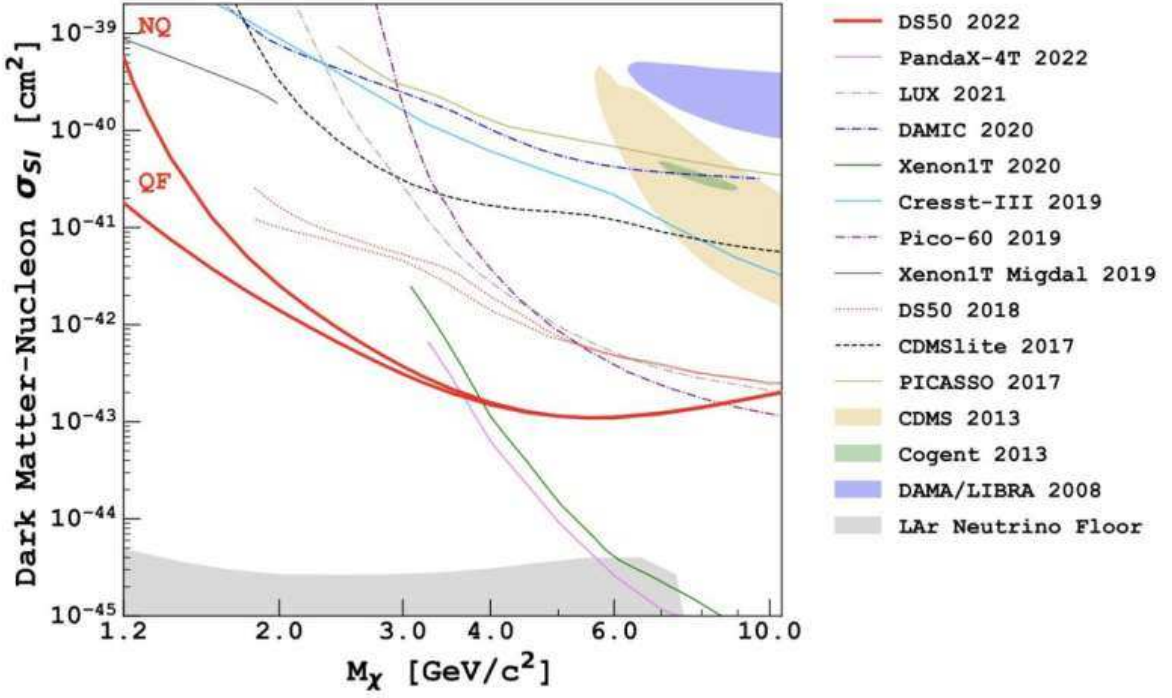}
 	\caption{\label{comp} Upper limits on the dark matter nucleon cross section as a function of the dark matter mass from various direct detection experiments. The lilac region in the top right of the plot corresponds to the DAMA data interpreted as due to WIMPS, with a cross section of $\sigma = 10^{-40}cm^2$. However in the same mass region of order 10 GeV the limits from some of the xenon experiments go down to  
 	 $10^{-44}cm^2$ or $10^{-45}cm^2$, i.e. about 5 orders of magnitude in
 	contradiction with DAMA.}  
 	
 \end{figure}

 The really most mysterious result is that the xenon-based experiments do not
 ``see'' the dark matter, while DAMA ``sees'' it\footnote{We note that the ANAIS experiment has failed to see an annual modulation with
 NAI(Tl) scintillators and their results \cite{ANAIS} are incompatible with the DAMA-LIBRA results at 3$\sigma$ .}. We want to explain the
 seemingly severe contradiction on Figure \ref{comp} of 5 orders of
 magnitude say: the DAMA-LIBRA experiment see per kg roughly $10^5$ times
 more than the upper limits for the best xenon experiments. Compared to LUX it
 may only be a factor $10^3$ say, but even that is a severe contradiction.

 Our explanation should be that our dark matter bubble or pearl falls
 through the xenon experiment containers with liquid xenon essentially by free fall while they move much slower in the solid materials, especially the sodium
 iodide NaI. But in order to avoid the seasonal variation being smeared away, the pearls
 have to come the 1400 m down to the experiments in less than a year. So the
 velocity on the average down through the mountain has to be at least
 \begin{eqnarray}
 	\hbox{Sinking velocity } v_{sink} &\ge& \frac{1400m}{year}\\
 	&=& \frac{1400 m}{31556926 s}\\
 	&=& 4.4 *10^{-5}m/s.
 \end{eqnarray}
 This is to be compared with the velocity, which a pearl of dark matter
 achieves in free fall inside the tank with the fluid xenon making up the
 typical xenon experiment. Since the acceleration at the Earth's surface
 is $9.8 m/s^2$, the velocity reached in a tank of depth $d$ is given as
 \begin{eqnarray}
 	v_{fall}&\approx& \sqrt{d*9.8m/s^2}\\
 	&=& 3m/s \qquad \hbox{(for e.g $d=1m$ big tank)}
 \end{eqnarray}
 This velocity in the essentially free fall in the fluid can reach up
 to be of order $10^5$ times larger than the
 $4.4 *10^{-5}m/s$ needed to ensure the pearls come down within a year.
 
 It must though be said here that if we want the pearls to pass through the shielding somewhat faster than in a  year then one should expect
 to see something soon in the xenon experiments. Now, however, what the xenon
 experiments looked for was a nuclear collision, and they at first sorted away the electron events. Their limits for electron events are much less strong and
 thus there is indeed no contradiction as yet, provided that we believe the
 story of pearls being stopped by the shielding and released again when coming to the fluid.
 
 This means that indeed the slowed down particle idea could solve the otherwise
 extremely difficult problem of resolving the apparent contradiction between the DAMA-LIBRA experiment and the xenon ones!
 
 In article \cite{XenT} the XenonnT collaboration rechecked the
 ``electron recoil excess'' previously observed in the Xenon1T experiment, and found that there was no effect
 signalling anything strange any more. Compared to the old results \cite{Xe1T} we can roughly say that the number of events in one bin went down so that a previous excess of
 $ 30\, Events/(t*y*keV)$ becomes an upper limit of say $ 10\, Events/(t*y*keV)$
 and another bin went even less down. With the original - now to be considered
 wrong - measurement we found \cite{Corfu21} that there was a need for 
 the faster passage through the liquid xenon than through the DAMA NaI by a factor 250 in order to diminish the rate of electron observation per kg.  In other words we needed that the pearls of dark matter spend per kg
 a factor 250 shorter time in the liquid xenon than in the NaI. Now we see that
 this factor 250 is not enough but should be increased by a factor
 corresponding to the decrease from the  $ 30\, Events/(t*y*keV)$ to the 
 $10\, Events/(t*y*keV)$ in one bin. Already the next
 most important bin is much less significant, and we shall crudely estimate
 that adding the two bins would mean the effect went down by a factor of 4 just to be on what is now the upper limit. So the upper limit on the signal in XenonnT is $4*250 =1000$ times smaller than the signal in the DAMA-LIBRA
 experiment.
 
 If we say, for illustration, that the free fall velocity in the
 xenon experiments is 3m/s, then the speed through the NaI in DAMA-LIBRA needs to be 1000 times slower, so that the time spent in a region of a kg in DAMA-LIBRA becomes 1000 or more times longer than the corresponding time spent in a region of a kg in the liquid xenon experiments. In order to achieve this the velocity through the NaI or similar solids should be
 $\frac{3m/s}{1000} = 3 mm/s$. In this case the time taken to pass down through the earth shielding
 will be
 \begin{eqnarray}
 	\hbox{ ``Passage time''}_{on \; limit } &\approx & \frac{1400m}{3mm/s}\\
 	&=& 4.2*10^6s = \frac{4.2*10^6s}{3.1 *10^7y/s}\\
 	&=& 1.3 *10^{-1}y\\
 	&=& 1.5\, month
 \end{eqnarray}
 
 So our model is a bit in tension with the DAMA results. Namely it suggests 
 that the time of year for the maximal signal 
 from the dark matter observation in DAMA should be shifted to be a bit later
 - by 1.5 month - than the time of year estimated from the astronomical
 expectation of the motion of the rest system of the dark matter in the
 neighbourhood of our solar system.


\section{How much available dust?} 

 We would like to get an estimate from the various types of matter available
 in the Universe as to how much dust can be afforded to settle on the dark matter
 bubbles. This will provide a check of our ideas by comparing with the
 inverse darkness ratio $\frac{\sigma}{M}$ in the limit of low velocity, where
 one would expect the dust to play the dominant role in the case that
 the dust grain is larger than the proper bubble.
 
 Let us begin by reviewing the amounts of matter of the different types available in the Universe:

  \begin{itemize}
  \item  27\% dark matter (while  68\% of a form of energy known as dark
    energy, and 5 \% ordinary matter).

\item The elements heavier than hydrogen and helium make up of order 2$\%$ of ordinary matter and are known as "metals''. Oxygen is the most abundant "metal" making up about $1\%$, carbon about $0.5\%$ and iron down to $0.1\%$ The comoving density of these "metals'' together \cite{density} is
\begin{eqnarray}
 \hbox{"metal'' density} &\approxeq& 5.0*10^6M_{\odot}Mpc^{-3}\\
 &=& 3.7*10^{-31}kg/m^3.
 \end{eqnarray}
\end{itemize}
\subsection{Inverse Darkness $\frac{\sigma}{M}$}
Using the atomic radii we can calculate the cross sections for the following atoms:
\begin{eqnarray}
  \hbox{Hydrogen H: } r_H &=& 25 pm\Rightarrow \sigma_{H}=\pi r_H^2 =1963 pm^2 \\
  \hbox{Helium He: } r_{He} &=& 30 pm\Rightarrow \sigma_{He}=\pi*r_{He}^2=
  2827pm^2\\
  \hbox{Carbon C: } r_C &=& 70 pm \Rightarrow \sigma_C=\pi*r_C^2=15394pm^2 \\
  \hbox{ Silicon Si: } r_{Si} &=& 110 pm \Rightarrow \sigma_{Si}=\pi*r_{Si}^2=
  38013 pm^2
  \end{eqnarray}

Using that one atomic unit $1 u$ = $1.66*10^{-27}kg$ we get for the inverse
darkness ratios $\frac{\sigma}{M}$ for the atoms mentioned:

\begin{eqnarray}
  \hbox{Hydrogen H: } \frac{\sigma_H}{1u*1.66*10^{-27}kg/u}&=& 1.18*10^6m^2/kg\\
  \hbox{Helium He: }\frac{\sigma_{He}}{4u*1.66*10^{-27}kg/u}&=& 4.26*10^5m^2/kg\\
  \hbox{Carbon C: }\frac{\sigma_C}{12u*1.66*10^{-27}kg/u}&=&7.73 *10^5m^2/kg\\
  \hbox{Silicon Si: }\frac{\sigma_{Si}}{28u*1.66*10^{-27}kg/u}&=& 8.18*10^5m^2/kg
  \end{eqnarray}


In a dust grain the atoms will typically shadow each other and thus this
ratio ``the inverse darkness'' will be smaller than if the atoms were all
exposed to the collision considered. If we denote the average
number of atoms lying in the shadow of one atom by ``numberthickness'' we
will have for the ratio for the full grain say
\begin{eqnarray}
  \frac{\sigma}{M}|_{grain}&=& \frac{\frac{\sigma}{M}|_{atom}}
       {\hbox{``numberthickness''}} 
       \label{shadow}
\end{eqnarray}
If the dust grain has a fractal dimension 2 or less there is no shadowing and the parameter "numberthickness'' = 1. 

  
If we insert in the grain a mass-wise dominating bubble, the whole object
will of course get a smaller ratio due to the higher mass,
\begin{eqnarray}
  \frac{\sigma}{M}|_{composed} &=& \frac{\sigma}{M}|_{grain}*\frac{M_{grain}}{M},
  \label{a}
\end{eqnarray}
where $M$ is the mass of the bubble or, if it dominates the whole composite
object, the dark matter particle.


On the average of course the mass ratio $\frac{M_{grain}}{M}$ of the dust
around the bubble and the bubble itself can never be bigger than the ratio of the amount of dust-suitable mass to that of the dark matter in the Universe. So
noting that the grain should largely be made up of the elements
heavier than helium, the so-called ``metals'', and that these make up
only of the order of 1\% of the ordinary matter which again is only
about 1/6 of the mass of the dark matter, we must have
\begin{eqnarray}
  \frac{M_{grain}}{M}&\le & 1\% /6 = 1/600.
  \label{b}
  \end{eqnarray}
But really of course not all the ``metal'' has even reached out to the
intergalactic medium, let alone been caught up by the
dark matter. So we expect an appreciably smaller value for this
ratio of dust caught by dark matter relative to the dark matter itself.
Taking $7*10^5 m^2/kg$ as the typical $\frac{\sigma}{M}|_{atom} $ for dust-suitable atoms and using the so to speak simplest ``equilibrium'' value for
$\frac{M_{grain}}{M}$ by combining equations (\ref{a}, \ref{b}) and using
$\hbox{``numberthickness''} =1$, we get
\begin{eqnarray}
	\hbox{First estimate: } \qquad \frac{\sigma}{M} &\approx& \frac{7*10^5m^2/kg}{600}\\
	&=& 1.2*10^3 m^2/kg = 1200m^2/kg.
\end{eqnarray}
  
In earlier papers we have already used the dark matter self-interaction in the low velocity limit extracted from Correa's fit to the dwarf galaxy data shown in Figure \ref{Correa} to give 
\begin{eqnarray}
  \frac{\sigma}{M}|_{v--> 0} &=& 15 m^2/kg.
\end{eqnarray}
So our first estimate comes out about a factor 100 over the observed value. However the dust on the dark matter was collected in a much earlier era than today, at least at first. 

We now wish to use a crude estimate of the amount of dust that might pile up around a dark matter bubble with a given velocity during the evolution of the Universe. There are two important effects to be taken into account. First of all the metal density was higher in the past due to the reduction in the ``radius" of the Universe by a factor $(1+z)^{-1}$ where $z$ is the red shift. Secondly the metallicity was lower in the past and we use the linear fits of De Cia et al. \cite{met} to its z dependence in our estimate of the rate of collection of metals by our pearls. 
We hence find that the most important time for the rate of collection of metals corresponds to $z = 3.3$, when the age of the Universe was 1.52 milliard years. At this time the rate of collecting metals for a given velocity was about 8.4 times bigger than if using the present metallicity and density. So we might crudely estimate the amount of dust then being collected by using an {\em effective metal density} of
\begin{eqnarray}
 \hbox{``metal density''}_{eff} &=&  3.71*10^{-31}kg/m^3*8.4\\
&=& 3.1*10^{-30}kg/m^3\\
&=&1.7*10^{-3}GeV/c^2/m^3.	
\end{eqnarray}

The metallicity at that time \cite{met} was a factor $10^{-1}$
times the one today. So the factor $1/600$ in equation (\ref{b}) for
the  ``metals'' accessible to be caught by the composite
dark matter particle becomes
\begin{eqnarray}
	\frac{M_{grain}}{M} &=& 1\%/6 /10 = \frac{1}{6000}.
	\label{available}
\end{eqnarray}
So taking $\frac{\sigma}{M}|_{atom} = 7*10^5m^2/kg$ for the atoms of dust and assuming all the accessible "metals'' are caught by our dark matter pearls,
we obtain our estimate for the inverse darkness of the dark matter particle
composed with a dust grain of dimension 2 or less:
\begin{eqnarray}
	\frac{\sigma}{M}|_{composed} &=&
	\frac{\sigma}{M}|_{grain}*\frac{M_{grain}}{M}\\
	&=& 7*10^5m^2/kg /6000 \\
	&=& 1.2*10^2m^2/kg.
\end{eqnarray}
This is of course really an upper limit for the inverse darkness, as not all of the "metal'' will have been caught up by the dark matter.

 Our estimated ratio
\begin{eqnarray}
	\frac{\sigma}{M}|_{composed} &=& 120 m^2/kg,
\end{eqnarray}
is actually one order of magnitude larger than the value extracted from the dwarf galaxy data
\begin{eqnarray}
	\frac{\sigma}{M}|_{Correa, v\rightarrow 0}&=&
	15 m^2/kg. 
\end{eqnarray}
This can be considered as a success for our model.

We now consider whether there is enough time to collect up so much dust.
For orientation we could first ask how much metal-matter at all could be
collected by a dust grain while already of the order of $10^{-7}m$ in size, meaning a cross section of $10^{-14}m^2$, and with a velocity of say 300 km/s =
$3*10^5 m/s$ during an effective age of the universe of
$1.52\, milliard$ years = 
$4.8*10^{16}s$. We obtain
\begin{eqnarray}
  \hbox{``available ``metals'' ''}_{for 10^{-14}m^2, 300km/s}&=&
  3*10^5m/s *4.8*10^{16}s*10^{-14}m^2*3.1*10^{-30}kg/m^3\nonumber\\
  &=&4.4*10^{-22}kg \\
  &=&2.4*10^5 GeV,
  \end{eqnarray}
which is to be compared to what the mass of a $(10^7m)^3$ large dust particle
with say specific weight $1000kg/m^3$ would be, namely
$10^{-18}kg$. So such a ``normal'' size $0.1 \mu m$ dust grain could not collect itself in the average conditions in the Universe.



However, if the grain to be constructed had lower dimension than 3, it would
help because then the cross section could be larger for the same hoped
for volume and thus mass. Decreasing say the thickness in one of the dimensions
from the $10^{-7}m$ to atomic size $10^{-10}m$ would for the same
collection of matter give a 1000 times smaller mass. This
would bring such a ``normal size'' grain close to being just collectable in the average conditions in the Universe.

Our speculated stronger forces than usual due to the big homolumo gap
would not help much, because the grain cannot catch the atoms in intergalactic space which it does not come near enough to touch.

  

  \subsection{Size of Individual Dark Matter Particles}
  
   In the approximation of dark matter having only a gravitational interaction it is  well-known that only the {\em mass density} matters, whereas the number density or the {\em mass per particle is not observable.}
  
  With other than gravitational interactions one could hope that it
  would be possible to extract from the fits in say our model, what the
  particle size should be. But the possibility
  for that in our model is remarkably bad! The Correa measurement yields just the ``inverse
  darkness'' ratio
  \begin{eqnarray}
  	\frac{\sigma}{M}&=& \frac{\hbox{``cross section''}}{\hbox{mass}} 
  \end{eqnarray}
  
  Our earlier very crude estimate \cite{Corfu21} for the rate of 3.5 keV radiation from dark matter seen by DAMA was based on:
  \begin{itemize}
  	\item The total kinetic energy of the dark matter hitting the Earth
  	per $m^2$ per $s$ (but not on how many particles).
  	\item The main part of that energy goes into 3.5 keV radiation of
  	electrons.
  	\item Estimate of a ``suppression'' factor for how small a part of this
  	electron radiation comes from sufficiently long living excitations to
  	survive down to 1400 m into the Earth.
  \end{itemize}
 None of this depends in our estimate on the size of the dark matter particles (provided it lies inside
a very broad range)!


If the dark matter particles were so heavy that the number density is so low
that the observation over an area of about $1 m^2$ would not get an event through
every year, then it would contradict the DAMA data.
The rate of dark matter mass hitting a square meter of the Earth is
\begin{eqnarray}
	\hbox{Rate} &=& 300km/s *0.3 GeV/cm^3\\
	&=& 3*10^5m/s *5.34*10^{-22}kg/m^3\\
	&=& 1.6*10^{-16}kg/m^2/s\\
	&=&5*10^{-9}kg/m^2/y
\end{eqnarray}
Taking the DAMA area of observation $\sim 1m^2$ we need to get more than
one passage per year and thus
\begin{eqnarray}
	M &\le & 5*10^{-9}kg\\
	&=& 3*10^{18}GeV.
\end{eqnarray}
Using the bubble internal mass density as estimated from the 3.5 keV homolumo gap \cite{Corfu21}, this upper bound implies that  
the bubble radius $R\le 10^{-7}m$.

\begin{figure}
	\includegraphics[scale=0.7]{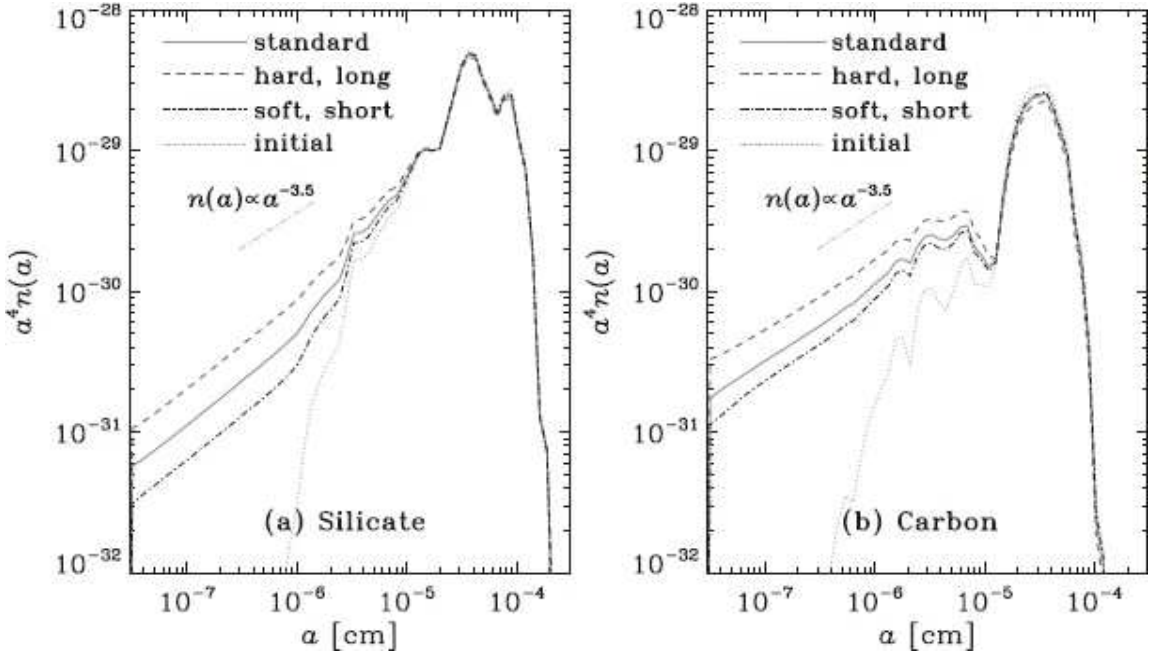}
	\caption{Simulated size distribution for dust grains.}  
	\label{size}
\end{figure}

If the bubble is small compared to the surrounding dust grain, then we would
not expect the development of the dust grain around the dark matter bubble
to be so different from the dust grains out in space not having any dark
matter bubble in them.
In Figure \ref{size}  we present the size distribution for dust grains from a simulation for ordinary dust particles of small dust grain production in galaxies \cite{distribution}. We take a typical grain size of $10^{-7}m$ say. Then using the low velocity limit $\frac{\sigma}{M} =15 m^2/kg$
gives
\begin{eqnarray}
	M&=& (10^{-7}m)^2/(15 m^2/kg)\\
	&=& 7*10^{-16}kg.
\end{eqnarray}
Such a 3 dimensional compact grain of dust has "numberthickness'' $\simeq$ 1000 atoms behind each other.
But if now the dust grain is 2 or lower dimensional, but keeping the mass of the compact grain combined with  $\frac{\sigma}{M} =15 m^2/kg$, there is no shadowing and
we obtain a 1000 times bigger estimate for the total or say bubble mass:
 \begin{eqnarray}
	M&=& 1000*7*10^{-16}kg = 7*10^{-13}kg 
\end{eqnarray}  
(for low dimensional dust of $10^{-7}m $ size).





\section{Summary Tables}
In Table \ref{suc} we summarize the successes of our model and give a brief explanation of each item below.


	\begin{table}
		\caption{\label{suc} Successes}
	\begin{adjustbox}{width=\columnwidth,center}
		\begin{tabular}{|c|c|c|c|c|
			}
			\hline
			\# \& exp/th & Quantity &value  &related Q. & value
			\\
			\hline
			1. & Dwarf Galaxies &&&\\
			exp & inv. darkness =& $15m^2/kg$& $\frac{M_{grain}}{M}$& $2*10^{-5}$\\
			th&= $\frac{\sigma}{M}|_{v\rightarrow 0}$ &$120 m^2/kg$ &&$1.6*10^{-4}$\\
			\hline
			2.& Dwarf Galaxies&&&
			\\
			exp &Velocity par. $v_0$ & $220km/s$
			& $4r_{dust}E$ &
			$8.1*10^{13}kg/s^2$
			\\
			th. & with hardening &$77 km/s$&$4r_{dust}E$    &$1*10^{13}kg/s^2$
			\\
			th. & without hard.& $0.7 cm/s$&$4r_{dust}E$&$400 kg/s^2$
			\\
			\hline
			3.& DAMA-LIBRA&&&
			\\
			exp &
			&$0.041 cpd/kg$ &suppression
			&
			$1.6*10^{-10}$
			\\
			th& air
			&$
			0.16 cpd/kg$&&$
			6*10^{-10}$
			\\
			th&stone & $1.6*10^{-5}cpd/kg$ &&$6*10^{-14}$
			\\
			\hline



		4.&Jeltema \&
		P.&&&
		\\
		exp&counting rate&$2.2 *10^{-5}phs/cm^2/s$&$\frac{\sigma}{M}|_{Tycho}$
		&$5.6*10^{-3}cm^2/kg$
		\\
		th&&$3*10^{-6}phs/s/cm^2$&$1\% *\alpha*\frac{\sigma}{M}|_{nuclear}$
		&$8*10^{-4}cm^2/kg$
		\\
		\hline
		5.&Intensity 3.5 kev&&&
		\\
		exp& $\frac{N\sigma}{M^2}$ & $10^{23}cm^2/kg^2$
		&$\frac{\xi^{1/4}}{\Delta V}$&$0.6 MeV^{-1}$
		\\
		th& & $3.6*10^{22}cm^2/kg^2$&&$0.5 MeV^{-1}$
		\\
		\hline
		6.& Two Energies&&&
		\\
		ast& line & 3.5 keV&&
		\\
		DAMA& av. en.& 3.4 keV&&
		\\
		\hline
	\end{tabular}
\end{adjustbox}
\end{table}



\begin{itemize}
\item{\color{blue} 1. Dwarf Galaxies, inverse darkness: } Assuming the dust grain around the
second vacuum bubble has a Hausdorff dimension of 2 or less and the ratio
of the dust grain mass to that of the whole dark matter particle to be
given by the total amount of metals in the gases in  the
space available  relative
to dark matter, we obtain (see equation \ref{available})
\begin{eqnarray}
	\frac{M_{grain}}{M}&=& 1.6*10^{-4}.
\end{eqnarray}
This leads to our estimate of the low velocity inverse darkness for our pearls
\begin{eqnarray}
	 \frac{\sigma}{M}|_{v=0,th} &=&
	120 m^2/kg,
	\end{eqnarray}
	and should be compared to Correa's value from her analysis of dwarf galaxies \cite{CAC}
	\begin{eqnarray}
    \frac{\sigma}{M}|_{v=0, ex}&=& 15 m^2/kg,\\
	\hbox{which would correspond to}\qquad \frac{M_{grain}}{M}|_{fitted} &=&2*10^{-5}. 
\end{eqnarray}
\item{\color{blue}2.  Dwarf galaxies, velocity parameter: }We seek to estimate
the velocity scale at which the inverse darkness $\frac{\sigma}{M}$
as a function of velocity falls significantly. One seeks to estimate the
velocity $v_0$ at which the colliding pearls pass so undisturbed through each other that the bending of their paths would not disturb the motion of the pearls enough to
influence the effective interaction seen by say Correa. The first
theoretical line marked ``with hardening'' has assumed that the dust around the pearl has been influenced by the exceptionally large homolumo gap in the
bubble to become extremely much harder than normal dust. It therefore
keeps a large cross section up to a much higher velocity,  $77 km/s$. The second theoretical line has just used a more normal elastic modulus for the dust, and even at rather small velocities, $0.7 cm/s$ and higher, the dust is
just deformed and makes
no effective bending of the motion of the pearl.
\item{\color{blue} 3. DAMA-LIBRA} We estimate crudely the amount
of kinetic energy in the dark matter pearls being converted, assumed mostly into the 
excitation of the 3.5 keV electron mode, during the stopping of the
pearls in the atmosphere ``air'' or in the solid part of the earth
``stone''. Next one asks how much there would be per kg of matter in the
earth if this energy were smoothly distributed over a range of depths
large enough to include the 1400 m deep Gran Sasso laboratory. The amount
of energy observed relative to this estimated available amount is denoted
``suppression'' and taken to be small because there are several different
excitations (although with mainly 3.5 kev) of different life times. However
the long life time ones require a longer time to be excited,
so that the ``suppression'' crudely becomes the ratio of the excitation
time available to the passage time down to the measurement place.
\item{\color{blue} 
	4. Jeltema and Profumo} This item estimates the amount of 3.5 keV
X-rays from the Tycho supernova remnant, from which it was rather
surprisingly observed by Jeltema and Profumo. We estimate the amount of
energy in the supernova remnant e.g. in the form of cosmic rays and
then believing that part of it hits the dark matter pearls and mainly
goes into emission of electrons. However a fraction of order the fine structure constant $\alpha$ of the excitation energy is emitted as 
X-rays, which like the electrons is supposed
to be mainly of the preferred energy per particle 3.5 keV.
We give a more detailed discussion of the Jeltma and Profumo results in  reference \cite{Bled20}.

After the workshop we decided that it is probably collisions with {\em hot atoms} rather than
cosmic rays that are predominant in making the dark matter produce the 3.5 keV radiation. This is
because the cosmic rays would stay around so long as to
make star formation regions like the centre of our galaxy produce more
3.5 kev X-rays than are seen.

\item{\color{blue} 
	5. Intensity 3.5 keV} 
	In the model picture, that the bubbles
of dark matter collide with one another and unite under a common skin, which contracts and thereby delivers energy of which an appreciable part is emitted as 3.5 keV X-rays, we estimate the overall intensity
of the expected 3.5 keV radiation. Cline and Frey \cite{Cline} have already fitted the observed
intensity from various astronomical objects of this 3.5 keV line as being
proportional to the 
{\em square}
$D^2$ of the density of dark matter $D$, from which one can extract
what in our model is $\frac{N\sigma}{M^2}$. Here $N$ is the number of
3.5 keV photons expected in one dark matter dark matter collision,
$\sigma$ the cross section, and $M$ the mass of a dark matter particle.
It turns out that our intensity quantity $\frac{N\sigma}{M^2}$ and the
very energy per photon (3.5 keV) are both obtained
as functions of the same parameter $\frac{\xi^{1/4}}{\Delta V}$ -
essentially the Fermi momentum of the electrons in the bubble. So
indeed the two quantities ``intensity'' and ``the 3.5 keV'' are related and the values of the parameter $\frac{\xi^{1/4}}{\Delta V}$ given in column 5 for the two quantities turn
out to agree. $\Delta V$ is the binding energy of a nucleon into the
inside-phase of the vacuum, and $\xi$ is the typical size of a dark matter
pearl relative to the ``critical size'' below which it would collapse.
\item{\color{blue} 
	6.  Two Energies} By this item we call attention to the
remarkable fact that the 3.5 keV X-ray line observed in astronomy, presumably
from dark matter, has the same energy per particle as the average energy of the
events in the DAMA-LIBRA experiment having the seasonal variation signalling
that they are from dark matter.


\end{itemize}





 \section{Addendum Since Workshop Held}
 \label{addendum}
 We would like to pre-announce some of the progress we have made since the workshop and which we hope to publish shortly:
 
 \begin{itemize}
 	\item We have had some success assuming that the 3.5 keV X-ray radiation
 	is not only produced in collisions between two dark matter particles, but
 	also when a dark matter pearl collides with an appropriate piece of  ordinary matter. Actually our speculation at present takes the direction
 	that, in the latter case, the major production of 3.5 keV X-rays should occur when atoms of
 	ordinary matter hit the dark matter. Then, namely, the atom can come with
 	more than 3.5 keV energy while the energy of the electron is still under
 	this value, so that the electron cannot penetrate into the bubble 
 	or even the hardened dust because its energy is below the homolumo gap
 	value. This should then mean that the electron would effectively be strongly interacting  with the dark matter pearl, while the atom still
 	has sufficient energy to excite the 3.5 keV line.
 	
 	The contribution of the dark matter ordinary matter interaction 
 	dominates in three cases:
 	\begin{itemize}
 		\item The observation of the outskirts of the Perseus Cluster, which we
 		previously dropped out of our fit because it produced a couple of
 		orders of
 		magnitude more 3.5 keV X-rays than estimated from the dark matter
 		density squared. One shall have in mind that the dark matter density
 		in the Perseus Cluster falls off with radius faster than the ordinary matter density, which is dominated by the so-called ``X-ray gas'', consisting of intra-cluster
 		hot gas or plasma revealing itself by its X-ray emission. This X-ray
 		gas is a very large component of the ordinary matter.
 		
 		\item The Center of the Milky Way: Actually Jeltema and Profumo \cite{Jeltema}
 		claim that the distribution of the 3.5 keV line from the Milky Way Center
 		does not look at all like coming from dark matter. If it reflects
 		the distribution of both dark matter and ordinary matter, the chance that it would vary in the
 		way found observationally should be better, since the ordinary matter
 		is distributed in a somewhat complicated way.
 		
 		\item Of course from the dark matter theory point of view the mysterious observation of the 3.5 keV line from the Tycho supernova remnant
 		can only be understood if some atoms of ordinary matter with X-ray temperatures are present in the
 		remnant and interact with the dark matter. 
 	\end{itemize}
 	
 	We found that the ratios of these three different observations of 3.5 keV line production can very crudely be compatible with a model of our type,
 	having both dark + dark production and dark + ordinary production.
 	
 \end{itemize}
 \subsection{Physics of the Vacuum Phases?}
 In 1998 Alford, Rajagopal and Wilczek proposed a new phase of QCD with
 ``Color-Flavour Locking'' \cite{Wilczek}. It was further argued that
 in matter with high baryon density \cite{Kryjevski, Alford} there should be
 a phase pattern with among other phases this color-flavour-locking one.
 In particular they found that,
for a density of the order of that in neutron
 stars, there was  a tetra-critical phase meeting point which in reality would
 probably be shifted into a couple of triple points. Remarkably this critical point occurred just for values of the strange quark mass and a common light quark mass (the heavy quarks being ignored) 
 close to the phenomenologically known ones! Provided this phase diagram
 could be extrapolated to zero baryon number density, we could claim that
 these authors have already found the phase(s) we hope for to produce our
 dark matter bubbles.

\begin{figure}
 	\vspace{-3cm}
 	\includegraphics[scale=0.8]{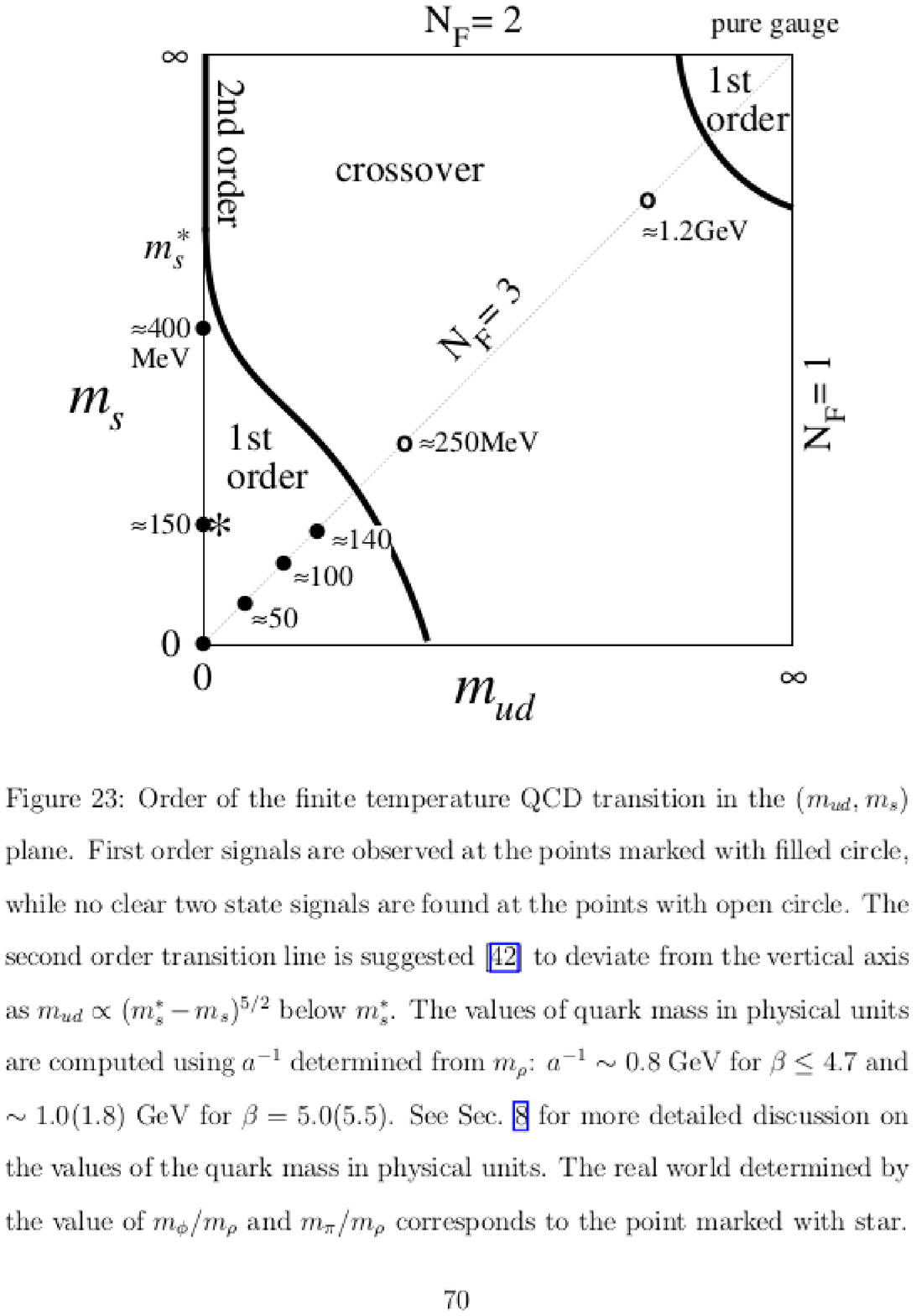}
 	\caption{\label{Columbia} This plot is taken from reference      \cite{Japanese}, which
 		used lattice calculations to find the phase transition
 		temperatures, and whether at this temperature there was a true phase
 		transition or just a crossover. The phase transition temperatures are
 		associated to the points in the quark mass plane at which the temperature was
 		found by their lattice calculations. The plot has the common light
 		quark mass on the abscissa and the strange quark mass on the ordinate.
 		They determined the physical point by looking for the correct meson masses,
 		and interestingly for us the $*$ denoting this physical quark mass combination is on the small quark mass side of the second order phase transition curve, contrary to the next figure \ref{threevar}.}
 \end{figure}

\begin{figure}
 	\includegraphics{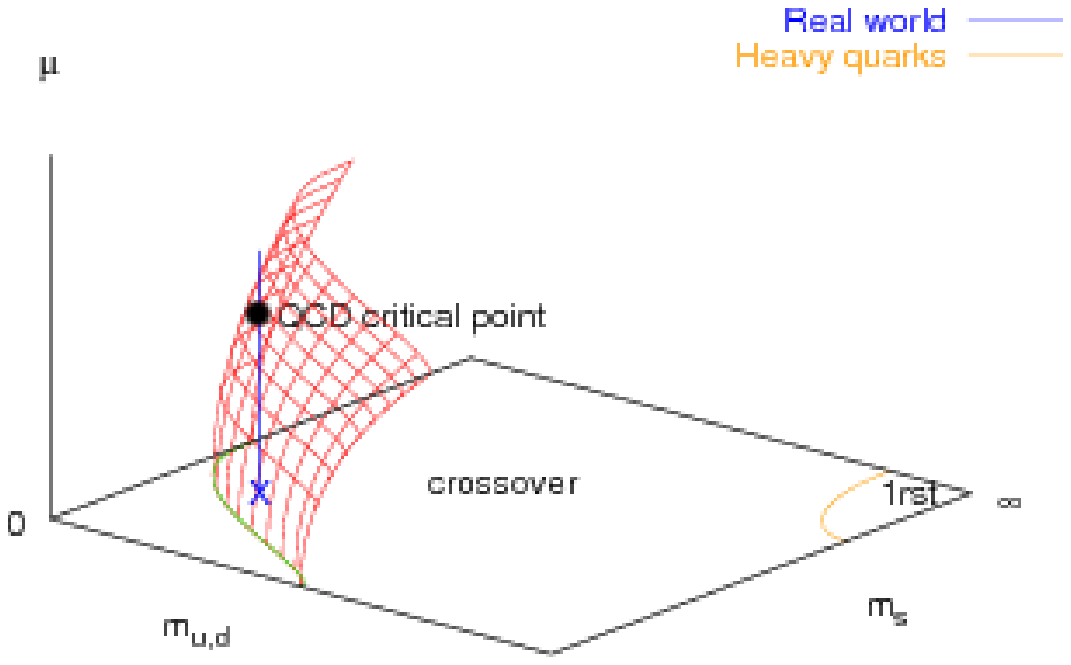}
 	\caption{\label{threevar} Here is a 3-dimensional phase diagram 
 		\cite{Columbiaplot} in which
 		there is imposed a chemical potential $\mu$ for baryon number along the
 		vertical axis, while at the bottom in perspective we see the usual
 		Columbia plot with the common u and d quark mass and the strange quark mass.
 		The figure is based on the result that the physical masses for the quarks are to the
 		large quark mass side of the second order phase transition. If however, as
 		we hope for, the physical quark mass combination would be exactly on the
 		second order  phase transition curve, then the QCD critical point on this figure would have moved down to the curve on the vacuum surface at $\mu =0$.}
 \end{figure}

 The phase diagram involving also the quark masses as parameters
 is the most important for us, because we would like to claim that
 Nature via  MPP would have fine-tuned the parameters, the quark
 masses in this case, to just let the vacuum be on the borderline among
 several phases. In the literature this phase diagram is known as the  Columbia-plot \cite{Cplot}, in which the abscissa is the mass, assumed to be the same, for the two light quarks $u$ and $d$,
 while the ordinate is the strange quark mass, the heavy quarks being ignored.
 The main property investigated in this plot is how raising the temperature can
 cause phase transitions. In the lower left corner of the plot one has
 spontaneous symmetry breaking of the chiral symmetry and a first order
 phase transition occurs as temperature is raised.
 For larger quark masses, 
 there is no true corresponding phase transition as the temperature is raised, but instead just a rapid variation of thermodynamical
 quantities. Such a situation is called a ``crossover''. It is our dream
 to realize our MPP by the experimental quark masses lying just on the curve
 separating off the mentioned lower left corner. This curve represents in the
 vacuum - i.e. zero temperature - a second order phase transition.
 The Columbia plot from  some time ago \cite{Japanese}, which we reproduce in 
 Figure \ref{Columbia}, has the physical point - meaning the experimental quark mass combination - inside the region in the lower left corner as limited off
 by the phase transition curve. However, Columbia plots from later lattice calculations mostly obtain this physical point lying in the region with the
 ``crossover'' as in Figure \ref{threevar} from reference \cite{Columbiaplot}. However an article using a technique of Ads/CFT correspondence type
 \cite{ColumbiaplotAdscft}
 also got the physical point inside the lower left corner region, in spite
 of the lattice calculations putting it on the bulk side
 (i.e. with the crossover). We are tempted to take this disagreement in the
 literature, as to even on which side of the (second order) phase transition
 curve the physical point lies, as saying that the uncertainties are
 such that one does not really know on what side this physical point lies.
 In fact that means then of course that it is, within the errors,
 possible that the physical point - as we dream - lies just on the
 phase transition curve.

 \subsubsection{What physics distinguishes the phases?}
 From the just mentioned hope for having a phase transition just at
 the physical quark mass values, we would come to think that the phases
 meeting at the experimental couplings are phases distinguished from each
 other just very little by the precise correlations between the Goldstone boson
 (the pions and the other pseudoscalar mesons) fields in the different
 phases. In fact we think one should imagine that in the phase in the lower left corner there is a significant correlation between the Goldstone
 fields at neighbouring places more than just that the fields are 
 lined up corresponding to the chiral symmetry breaking direction; while
 in the phase in the region marked ``crossover'' the correlations are closer
 to be only given by the quark masses. There will of course still be
 correlations in the "crossover'' phase, also between neighbouring regions, but they will be smaller
 correlations of that type than in the left lower corner phase.
 
 In later work we would hope to estimate the scale of distance at which 
 the Nambu-JonaLasinio spontaneous breakdown sets in, presumably it is
 order of magnitudewise given by $f_{\pi}=93 MeV$. Then we can understand that,
 when the Compton wavelengths of the pion and kaon are short compared to the
 scale for the spontaneous breakdown, the sigma model fields will be specified
 effectively alone by the quark masses at a short distance scale. When on the
 other hand the Compton wavelengths for the quarks are long, then in first approximation we have a true spontaneous breakdown and the quark masses only provide
 a weak force determining the direction of the breaking of the chiral symmetry.
 In the latter case we expect a first order phase transition to a situation
 where say due to higher temperature the spontaneous breakdown has disappeared.
 In the short Compton wavelength situation, however, there is no true
 spontaneous breaking, because there is all the time a state kept with the same
 orientation of the breaking even rather locally, and thus we only expect the
 crossover phase. As  $f_{\pi}$ and say the average mass for the three light
 quarks are not much different, our speculation of the phase transition
 along the second order curve just coinciding with the physical quark masses
 is not excluded.

\begin{figure}
 	\includegraphics[scale=0.8]{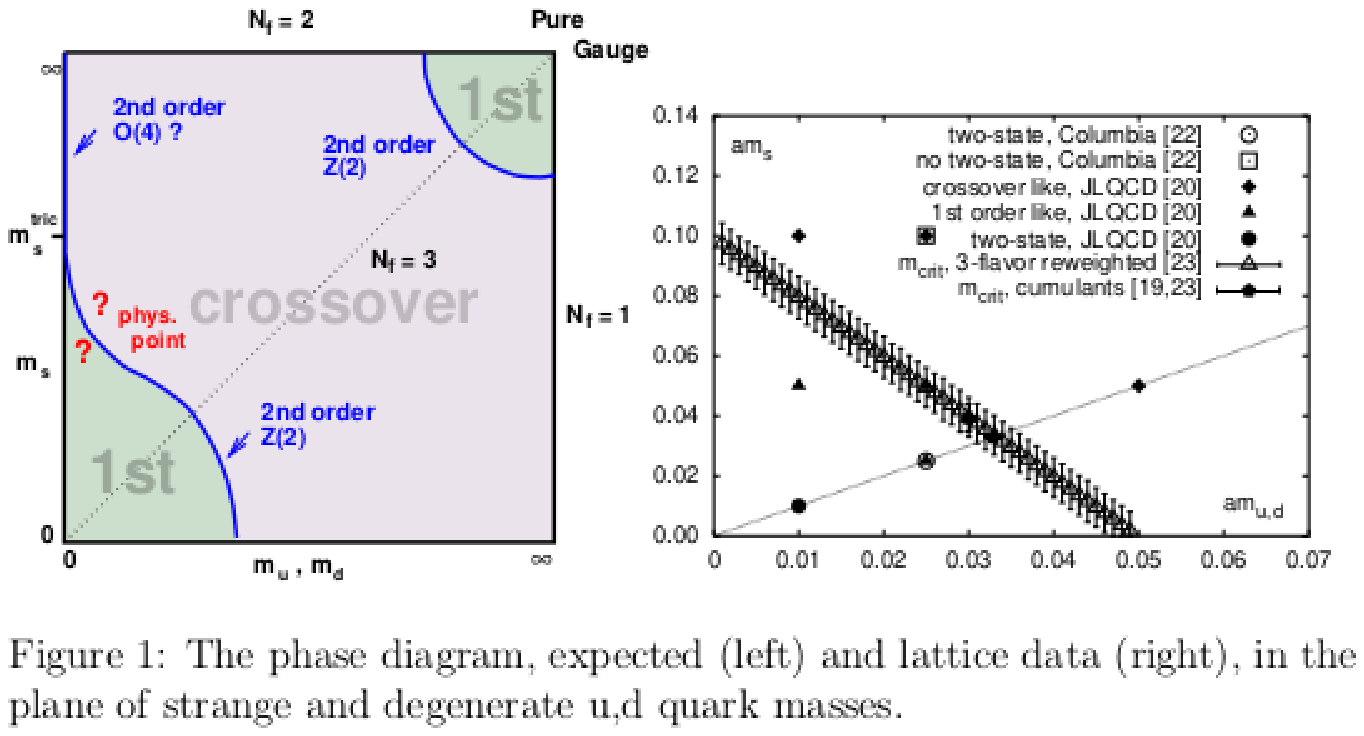}
 	\caption{\label{que} Here Learmann and Philipsen have left
 		question marks to tell that they did not know on what side of the second
 		order phase transition the physical combination of quark masses lie.
 		The right hand figure shows the second order phase transition determined from
 		their lattice simulations with the lattice constant multiplied by the
 		relevant quark mass along the axes.}
 \end{figure}

 In early work by Learmann and Philipsen \cite{question}, they directly admit that they do not
 know on what side of the second order phase transition the physical point
 lies by putting question marks on their Columbia plot; see Figure \ref{que}. 
 So we believe that it is not overly optimistic
 to suggest that some different phases in QCD really can exist if the
 quark masses just take the appropriate values, and that these values
 are indeed very close to the experimental quark masses.
 
 If this is how our speculated phases come about then the scale of this
 physics is surely the strong interaction scale, say 0.4 GeV or so.
 However, we should probably take into account
 the very small difference in the correlations between the different phases which we suggested above. We would then 
 expect that compared to a simple dimensionality
 argument - giving say 0.4 GeV as the scale - the tension $S$ will be
 exceptionally small. So instead of expecting the cubic root $S^{1/3}$ to
 be 0.4 GeV, we expect it to be somewhat smaller.
 
 \subsubsection{What Tension did we get in our Dark Matter Fitting?}
 
 Now partly because, for dimensional reasons, we must consider the cubic
 root $S^{1/3}$ and partly because the mass of course goes as the cube of the
 linear distance scale, the mass $M$ of our dark matter pearls depend on
 the ninth power of this cubic root $S^{1/3}$ of the tension. Thus putting
 in cubic roots of the tension only deviating by one order of magnitude
 we get masses deviating by 9 orders of magnitude. Thus our prediction of the
 mass $M$ from the order of magnitude estimate of $S^{1/3}$ taken to be
 of the strong interaction energy scale allows a rather large range of masses; for example
 \begin{eqnarray}
 	S^{1/3}= 1\,  GeV&\Rightarrow& M = 24\, kg\\
 	\hbox{while \quad }S^{1/3}= 100\, MeV &\Rightarrow& M=2.4*10^{-8}\, kg
 	\label{100MeV}\\
 	\hbox{and \quad }S^{1/3}=10\, MeV &\Rightarrow& M=2.4 *10^{-17}\,kg=1.4 *10^{10} \, u.
 \end{eqnarray}
 Here u $\sim$ GeV is the atomic unit.
 
 The $M=2.4*10^{-8}\, kg$ value would correspond to about 3000 km between the pearls in the solar
 system region, while $M=24\, kg$ would lead to a 1000 times longer distance,
 i.e. 3 million km between the pearls. To reach the mass required for
 making the DAMA experiment see enough separate particles, given above
 as the mass $M\le 5*10^{-9}kg = 5\, \mu g$, requires an $S^{1/3}$ value
 less than the 100 MeV in equation (\ref{100MeV}) by a factor of the ninth root of 4.8, which means 
 $84\, MeV$. This $84\, MeV$ is thus the upper limit to the ``strong interaction
 scale'' allowed for our model to work.
 However, as already suggested, the weakness of the phase transition probably leads to a somewhat low tension and reaching down below 84 MeV should be possible.
 
 \subsubsection{Getting Rid of Cosmological Constant and Replacing by Domain
 	Walls} \label{CC}
 In the article about our dark matter in the next workshop \cite{walls} on
 `` Tensions
 in Cosmology'', we calculate that one has the possibility of getting rid of
 the cosmological constant and effectively replacing it by domain walls.
 To achieve this modification of cosmology we need roughly
 \begin{eqnarray}
 	S^{1/3} &\sim & 30\, MeV,
 \end{eqnarray}
 which fits well as a diminished strong interaction scale less than the more conventional $0.4 $ GeV, as well as with the requirement of it being less than 84 MeV.
 
 This tension corresponds to a mass of our dark matter bubbles of $M \sim
 10^{-12}\, kg \sim 10^{15}\, GeV$.

 \section{Conclusion}
 
 We have again presented and developed our long-standing model for
 dark matter as macroscopic composites of ordinary matter kept together
 and concentrated by means of a new type of vacuum - another phase of the vacuum. We think that our model solves some problems that are very
 difficult for other models and, in addition, has some good order of magnitude
 agreements suggesting that our model could be on the right track.
 For example, the most mysterious fact that DAMA-LIBRA have seemingly seen the dark matter while several Xenon based experiments do not - even with significantly more sensitivity - see it. We suppose dark matter is something
 moving through the apparatuses with a speed quite different for a solid NaI
 detector as in DAMA-LIBRA, where the dark matter pearls can get stopped and thus spend more time in the apparatus, than in the fluid xenon apparatuses, where they
 will fall quickly through. Another achievement of our model is that,
 because it is so difficult  e.g. to see new vacua, we have essentially no
 new physics and can thus cope with the fact that LHC has seen
 nothing of the dark matter. Since the workshop we have even speculated about
 the so-called Columbia plots, which suggest some phase transitions
 as a function of the quark masses. If our speculations were established, we would really
 have the chance to avoid LHC having had to see anything concerning the dark
 matter - our model can so to speak live with {\em no new physics.}
 
 We now list a series of comments about our model:

 \begin{itemize}
 	\item We have put forward a model for dark matter consisting of
 	nm size bubbles of a {\em new vacuum} with a mass inside of $2*10^{-15} kg \sim 10^{12}\, GeV $ and, apart from this
 	vacuum speculation, using {\em only the Standard Model.}
 	(But recently we got the idea that, if one wants to change cosmology by
 	avoiding the dark energy and thus putting the cosmological constant to
 	zero, there is a chance of getting approximately the same effects from
 	domain walls on the borderlines of the different vacua. This requires
 	though that the pearls become bigger. In such a case the size of the
 	bubble should be of the order of $10^{-8}$ m, rather than the nm size we believed at the time of the workshop in Corfu 2022.)  
 	\item The low velocity interaction of the dark matter particles in outer space - especially in dwarf galaxies, where Correa estimated it - is given by a {\em dust grain} sitting around the pearl of the new
 	vacuum. Using the relative amount of ``metals'' to dark matter as also
 	giving the ratio of the masses of the grain of dust to the dark matter
 	in the dark matter particles, we get only an order of magnitude deviation
 	from the low energy ``inverse darkness'' estimated by Correa:
 	\begin{eqnarray}
 		\frac{\sigma}{M}_{v-->0} = 120 m^2/kg
 		&\hbox{against Correa: }& \frac{\sigma}{M}_{v-->0} = 15 m^2/kg.
 	\end{eqnarray}
 	
 	\item Because of having lesser darkness than ideal dark matter (which has
 	only gravitational interactions) - i.e. larger inverse darkness
 	$\frac{\sigma}{M} \sim 10^{-3}m^2/kg$, even after blowing off the attached dust grain, than the ``usual'' WIMPs - our dark matter particles get
 	slowed down in the atmosphere. This explains why the Xenon-experiments,
 	which only look for nuclei recoil events when expecting dark matter,
 	do not see any dark matter giving nuclei.
 	\item The highly compressed ``ordinary matter '' (in the
 	new vacuum) inside the bubble has, compared to usual chemistry, a very high energy gap - homolumo gap -
 	between the highest filled state, {\em HOMO}, and the lowest unoccupied
 	state, {\em LUMO}, speculated to be just $3.5\, keV$.
 	This leads to the dark matter bubble having a
 	``preferred'' {\em  emission energy for photons and electrons
 		given by the gap height 3.5 kev}.
 	
 	
 	\item It is this emission of 3.5 keV electrons (and photons) after
 	excitation of the ``ordinary'' material inside the bubble, which we suppose is
    observed by DAMA. 
 	
 	\item It is a remarkable accident supporting our model that the
 	direct observation at DAMA-LIBRA
 and
 the X-ray radiation observed from galaxy clusters and galaxies
 supposedly coming from dark matter
 both
 have the average energy per event of  3.5 keV.
 \end{itemize}
 \section{Acknowledgement}
 Both of us thank our institutes for our status as emeriti, and one of us
 in addition thanks the Niels Bohr Institute for economic support for the
 visit to the Corfu Workshop where this talk was presented, as well as
 for support to visit the Workshop in Bled which was also supported even with a
 contribution for printing and where one of the subjects was similar to this talk.
 



\begin{thebibliography}{99}
  \bibitem{DAMA1}
  R. Bernabei et al.,
  Eur. Phys. J. {\bf  C73}, 2648 (2013).
  [arXiv:1308.5109].
  	
  \bibitem{DAMA2}
  R. Bernabei et al,
  Prog. Part. Nucl. Phys. {\bf 114}, 103810 (2020).
  		
  \bibitem{Xe1T} E. Aprile et al., 
  Phys. Rev. {\bf D102}, 072004 (2020)
 [arXiv:2006.09721].
 
 \bibitem{XenT} E. Aprile et al.,
 Phys. Rev. Lett. {\bf 129}, 161805 (2022)
 [arXiv:2207.11330].
 
 \bibitem{CAC}
 C. A. Correa,
 MNRAS {\bf 503} 920 (2021)
 [arXiv:2007.02958].  
 
 \bibitem{Wilczek}
 M. Alford, K. Rajagopal and F. Wilczek,
 Nucl. Phys. {\bf B537}, 443 (1999)  
 [arXiv:hep-ph/9804403]. 
  
\bibitem{Cosinius}
G. Angloher et al.,
arXiv:2106.07390.

\bibitem{Dark1}
C.~D.~Froggatt and H.~B.~Nielsen,
Phys. Rev. Lett. {\bf 95} 231301 (2005)
[arXiv:astro-ph/0508513].

\bibitem{Dark2}	
C.D. Froggatt and H.B. Nielsen,
Proceedings of Conference: C05-07-19.3 (Bled 2005),
arXiv:astro-ph/0512454.

\bibitem{Tunguska}
  C.~D.~Froggatt and H.~B.~Nielsen,
  Int.\ J.\ Mod.\ Phys.\ A {\bf 30} no.13, 1550066 (2015)
 [arXiv:1403.7177].

\bibitem{supernova}
C. D. Froggatt and H. B. Nielsen,
Mod. Phys. Lett. {\bf A30} no.36, 1550195 (2015)
[arXiv:1503.01089].

\bibitem{Corfu17}
H.B. Nielsen, C.D. Froggatt and D. Jurman,
PoS(CORFU2017)075.

\bibitem{Corfu19}
H.B. Nielsen and C.D. Froggatt,
PoS(CORFU2019)049.

\bibitem{theline}
C. D. Froggatt, H. B. Nielsen,
``The 3.5 keV line from non-perturbative Standard Model dark matter balls'',
arXiv:2003.05018.

\bibitem{Bled20}
H. B. Nielsen (speaker) and C.D. Froggatt, ``Dark Matter Macroscopic Pearls,
3.5 keV -ray Line, How Big?'',
23rd Bled Workshop on What comes beyond the Standard Models (2020),
arXiv:2012.00445.
\bibitem{Bled21}
C. D. Froggatt and H.B.Nielsen, 
``Atomic Size Dark Matter Pearls, Electron Signal'',
24th Bled Workshop on What comes beyond the Standard Models (2021),
arXiv:2111.10879.
\bibitem{extension}
C. D. Froggatt and H.B. Nielsen, 
``Atomic Size Pearls being Dark Matter giving Electron Signal'',
arXiv:2203.02779.

\bibitem{Corfu21}
H. B. Nielsen and C. D. Froggatt,
PoS(CORFU2021) 095, 
 arXiv:2205.08871.

\bibitem{Bled22}
H. B. Nielsen and C. D. Froggatt
25th Bled Workshop on What comes beyond the Standard Models (2022),
arXiv:2303.06061.

\bibitem{walls} 
H. B. Nielsen and C. D. Froggatt,``Domain Walls of
Low Density in Cosmology'' 
in ``Tensions in
Cosmology'' in Corfu Summer Institute 2022.

\bibitem{MPP1}
D.~L.~Bennett, C.~D.~Froggatt and H.~B.~Nielsen,  NBI-HE-94-44, GUTPA-94-09-3,
Presented at Conference: C94-07-20 (ICHEP 1994), p.557-560.

\bibitem{MPP2}
D.~L.~Bennett, C.~D.~Froggatt and H.~B.~Nielsen,
NBI-95-15, GUTPA-95-04-1,
Presented at Conference: C94-09-13 (Adriatic Meeting 1994), p.255-279,
arXiv:hep-ph/9504294.


\bibitem{MPP3}
D.~L.~Bennett and H.~B.~Nielsen,
Int. J. Mod. Phys. {\bf A9} 5155 (1994).

\bibitem{MPP4}
D.~L.~Bennett, C.~D.~Froggatt and H.~B.~Nielsen,
NBI-HE-95-07,
Presented at Conference: C94-08-30 (Wendisch-Rietz) p.394-412.

\bibitem{Picek}
D. L. Bennett, H. B. Nielsen and I. Picek,
Phys. Lett. {\bf B208} 275 (1988).

\bibitem{book}
C. D. Froggatt and H. B. Nielsen,
Origin of Symmetries,
World Scientific Publishing Co. Pte. Ltd. (1991).  

\bibitem{tophiggs}
C.~D.~Froggatt and H.~B.~Nielsen, Phys. Lett. {\bf B368} 96 (1996)
[arXiv:hep-ph/9511371].

\bibitem{Corfu1995}
H.B. Nielsen (Speaker) and C.D. Froggatt,
Presented at Conference: C95-09-03.1
(Corfu 1995),
arXiv:hep-ph/9607375.

\bibitem{Cline}
J. M. Cline and A. R. Frey,
Phys. Rev. {\bf D90}, 123537 (2014)
[arXiv:1410.7766].

\bibitem{Jeltema}
T. Jeltema and S. Profumo,
MNRAS {\bf 450}, 2143 (2015)
[arXiv:1408.1699].

\bibitem{Hd}Ryan Hayes and Michael Freed, March 2007,
Journal of Young Investigators,
``The Fractal Dimension and Charging of Preplanetary Dust Aggregates''
https://www.jyi.org/2007-march/2017/11/11/the-fractal-dimension-and-charging-of-preplanetary-dust-aggregates

\bibitem{Wright}
E. L. Wright,
ApJ {\bf 320}, 818 (1987),
htpps://astro.ucla.edu/wright/dust/

\bibitem{NFW}
J. F. Navarro, C. S. Frenk and S. D. M. White, 
ApJ. {\bf 462}, 563 (1996)
[arXiv:astro-ph/9508025].

\bibitem{firstSIDM}
D. N. Spergel, and P. J. Steinhardt,
Phys. Rev. Lett. {\bf 84}, 3760 (2000) 
[arXiv:astro-ph/9909386].

\bibitem{ANAIS}
 J. Amar\'{e} et al., Phys. Rev. {\bf D103}, 102005 (2021) 
 [arXiv:2103.01175].

\bibitem{density}
F. Calura and F. Matteucci,
  Mon. Not. R. Astron. Soc. {\bf 350}, 351 (2004)
  [arXiv:astro-ph/0401462].


\bibitem{met}
A. De Cia et al.,
Astron. Astrophys. {\bf 611}, A76 (2018).
\bibitem{distribution}
H. Hirashita and H. Kobayashi,
Earth, Planets and Space {\bf 65}, 1083 (2013).

  \bibitem{Kryjevski}
  A. Kryjevski, D. Kaplan and T. Sch\"{a}fer,
  Phys. Rev. {\bf D71}, 034004 (2005).

\bibitem{Alford}
M. G. Alford et al.,
Rev. Mod. Phys. {\bf 80}, 1455 (2008)
[arXiv:0709.4635].

\bibitem{Cplot}
F. R. Brown et al.,
Phys. Rev. Lett. {\bf 65} 2491 (1990).

\bibitem{Japanese}
Y. Twasaki et al.,
arXiv: Hep-lat/9605030.


\bibitem{Columbiaplot}
P. de Forcrand and M. D'Elia,
arXiv:1702.00330.

\bibitem{ColumbiaplotAdscft}
S. Bartz  and T. Jacobson, 
	Phys. Rev. {\bf C97}, 044908 (2018)
[arXiv:1801.00358].


\bibitem{question}
E. Laermann and O. Philipsen,
Ann. Rev. Nucl. Part. Sci. {\bf 53}, 163 (2003)
[arXiv:hep-ph/0303042].





 \end{thebibliography}
\end{document}